\documentclass[a4paper,fleqn]{cas-sc}

\usepackage[authoryear,longnamesfirst]{natbib}
\usepackage{amsmath}
\usepackage{mathtools}
\usepackage{float}
\usepackage{placeins}
\usepackage{subcaption}
\usepackage{algorithm}
\usepackage{algpseudocode}
\usepackage{tabularx}
\usepackage{booktabs}
\usepackage{svg}
\usepackage{pdfpages}
\usepackage{graphicx}
\usepackage{bm}
\def\tsc#1{\csdef{#1}{\textsc{\lowercase{#1}}\xspace}}
\tsc{WGM}
\tsc{QE}
\tsc{EP}
\tsc{PMS}
\tsc{BEC}
\tsc{DE}

\begin{document}
\let\WriteBookmarks\relax
\def\floatpagepagefraction{1}
\def\textpagefraction{.001}
\shorttitle{}
\shortauthors{Neeraj Balachandar, A. Padmaprabhan, Vishnu R. Unni}

\title[mode = title]
{An Aeroelastic Solver Integrating Reformulated-Vortex-Particle and Finite-Element Methods across Non-Conforming Interfaces}


\author[1]{Neeraj Balachandar}[
    orcid=0009-0006-7210-9648
]
\fnmark[1]
\ead{neerajbalachandar@gmail.com}

\author[1]{A. Padmaprabhan}[
    orcid=0009-0004-0797-9546
]
\fnmark[1]
\ead{padmaprabhan2004@gmail.com}

\author[1]{Vishnu R. Unni}
\cormark[1]
\ead{vishnu.runni@mae.iith.ac.in}

\affiliation[1]{organization={Department of Mechanical and Aerospace Engineering, Indian Institute of Technology Hyderabad},
                addressline={Sangareddy}, 
                city={Hyderabad},
                postcode={502284}, 
                state={Telangana},
                country={India}}

\cortext[cor1]{Corresponding author}

\fntext[fn1]{These authors contributed equally to this work.}

\begin{abstract}
We present \texttt{VarFlExI} (\underline{Var}iable \underline{F}idelity Unsteady \underline{Fl}ow--\underline{F}EniCS \underline{Ex}change \underline{I}nterface), a modular aeroelastic solver for modeling two-way fluid–structure interaction (FSI) of flexible lifting surfaces. The framework employs a \textit{reformulated Vortex Particle Method (rVPM)} to solve the incompressible \textit{Navier-Stokes} equations without the need for computationally expensive volumetric meshing, while supporting variable-fidelity aerodynamic modeling using the \texttt{FLOWUnsteady} framework. 
On the structural side, a \textit{Reissner-Mindlin} plate formulation is discretized using the finite element method and integrated in time through the generalized-$\alpha$ method, with the nonlinear equilibrium equations solved using a Gauss-Newton procedure within the FEniCS framework. Fluid and structural solvers are coupled through an explicit staggered partitioned scheme, ensuring conservation of virtual work for load and displacement transfer across the non-matching interface. To accommodate the multi-representative nature of the aerodynamic loads and geometry, the interface coupling employs separate work-conservative force and reverse-geometry transfer operators via a common intermediate interface. The framework is validated against water-tunnel experiments, demonstrating accurate prediction of the coupled aeroelastic response rather than independent validation of the constituent solvers. The computational efficiency of the meshless aerodynamic solver enables simulations at significantly lower computational cost while maintaining accuracy. The solver is further evaluated through sensitivity analyses and parameter studies spanning different flow conditions, structural properties, and coupling parameters. 




 
\end{abstract}




\maketitle

\section{Introduction}

\begin{figure}[pos=h!]
    \centering
    \includegraphics[width=\linewidth]{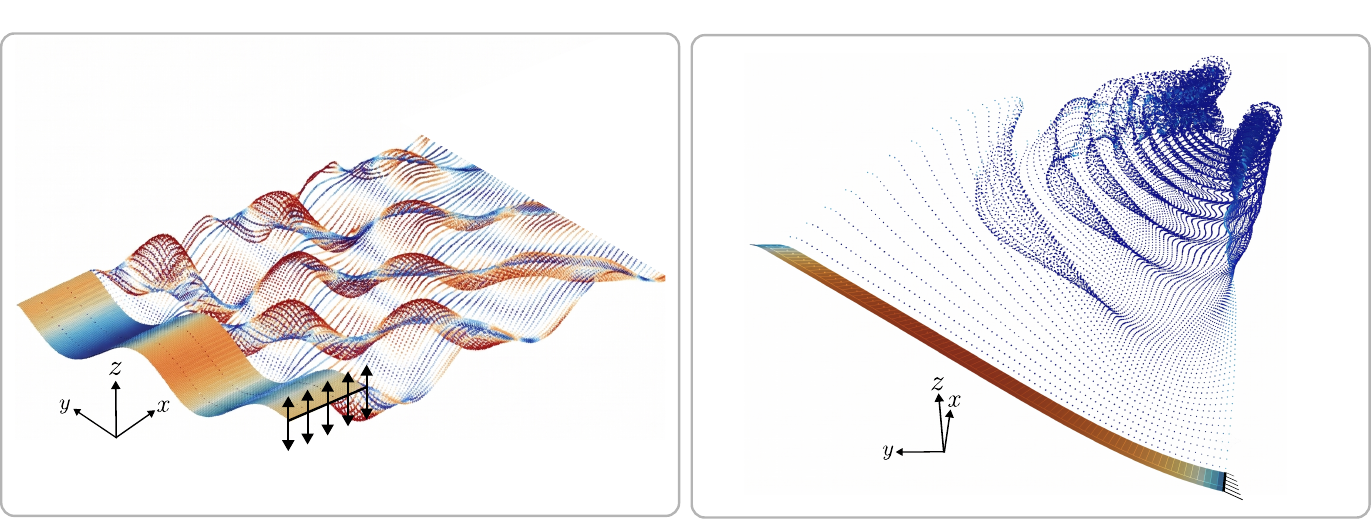}
    \caption{Schematic overview of \texttt{VarFlExI}, illustrating the two-way aeroelastic coupling between
    the \texttt{FEniCS}-based structural dynamics and the \texttt{FLOWUnsteady} aerodynamics. The flexible lifting surface is represented by a \textit{\textbf{Reissner-Mindlin}} structural model, while the aerodynamic field is represented by the \textit{\textbf{reformulated Vortex Particle Method}} (\textit{rVPM}). (\textbf{Left}) The simulation depicts a \textbf{highly flexible} (Elasticity modulus $\approx 10^8$), dynamically deforming wing undergoing \textbf{root-driven heaving} with pronounced spanwise bending-wave formation. (\textbf{Right}) Depicts a highly flexible wing undergoing \textbf{aeroelastic flutter}, with fixed root.}
    \label{fig:abstract}
\end{figure}

Aeroelasticity concerns the coupled interaction between aerodynamic loading and structural dynamics and plays a central role in the analysis and design of aerospace structures subjected to unsteady flow. The interaction becomes particularly important for lightweight and flexible lifting surfaces, for which aerodynamic forces can substantially modify the structural response, while structural deformation simultaneously alters the surrounding flow field. Such feedback can produce a range of static and dynamic aeroelastic phenomena, including dynamic response to gusts and maneuvers, flutter, and nonlinear limit-cycle oscillations \cite{chai_flutter}. Consequently, reliable prediction of aeroelastic stability and transient structural response is an important component of aircraft design, optimization, and performance evaluation \citep{raveh_intro}. Since standalone aerodynamic or structural analyses are insufficient for these flexible systems, modeling the behavior of wingtip vortices and the trailing wake is critical to understanding their aeroelastic stability \citep{kamakoti_intro, higharwing}.

In a conventional CFD formulation, the aerodynamic field is typically discretized using a fixed or deforming volumetric mesh, requiring structural deformations to be communicated back to the fluid solver at every time step. Consequently, large structural motions necessitate continuous mesh deformation and regeneration, while the coupled time integration demands expensive, iterative evaluations of both the fluid and structural systems. Classical aeroelastic analyses often rely on reduced aerodynamic representations, linearized unsteady aerodynamic theories, or frequency-domain methods. These approaches remain highly valuable due to their low computational cost and suitability for preliminary design and flutter-boundary estimation \citep{rom_intro1, rom_intro2}. However, they become overly restrictive when applied to complex aerospace and automotive problems. These advanced applications frequently involve dynamic aeroelasticity, highly unsteady flows and nonlinear aerodynamic effects, complex three-dimensional wake dynamics, and large, nonlinear structural deformations. Consequently, relying solely on high-fidelity, CFD-based simulations imposes prohibitive computational costs, especially during iterative design \citep{raveh_intro, shape_opt}. Transitioning to a reduced-order modeling framework overcomes this bottleneck, enabling rapid, control-oriented aeroelastic optimization \citep{control_aero_brunton}. Furthermore, these advanced reduced-order models pave the way for a novel class of differentiable solvers, introducing powerful gradient-based capabilities to coupled fluid-structure problems \citep{jax_aero_intro}.

These limitations motivate the use of aerodynamic models that provide an intermediate level of physical fidelity while retaining substantially lower computational cost. A broad range of aerodynamic models exists along this fidelity spectrum, from lifting-line and vortex-lattice approaches to panel methods, wake-based formulations, and full Navier–Stokes solvers \citep{unsteadyvlm}. The appropriate choice depends on the physical phenomena of interest and the required computational throughput. In particular, mid-fidelity methods based on vortex dynamics can retain important unsteady wake physics without requiring the full volumetric resolution of a conventional CFD solver. \textit{Vortex Particle Methods (VPM}) are particularly attractive in this context because the vorticity field is represented by Lagrangian particles, naturally providing a meshless description of the wake and avoiding many of the mesh generation and deformation requirements associated with Eulerian CFD \citep{vpm_intro_wind}. Vortex-particle-based aerodynamic models have been coupled with structural solvers for problems involving flexible structures, flutter, and transient aeroelastic response \citep{chau_harvester_vpm, vpm_appln_intro2}. The \textit{reformulated Vortex Particle Method (rVPM)} further provides a stable meshless large-eddy simulation formulation for three-dimensional incompressible flow, with computational efficiency that can be substantially higher than conventional mesh-based LES at comparable levels of aerodynamic fidelity \citep{alvarez2022flowunsteady}, hence making such approaches attractive for repeated unsteady aerodynamic evaluations involving vortex-dominated flows and for aeroelastic applications in which the aerodynamic solution must be recomputed continuously as the structure deforms. The utility of this intermediate-fidelity regime has also been demonstrated in coupled aeroelastic simulations \citep{aeroelasticmdo}.

Structural modeling constitutes a critical component of computational aeroelasticity, with its choice determining the accuracy, computational cost, and range of deformation that can be captured. Existing solvers employ structural models spanning from reduced-order beam and plate formulations to high-fidelity three-dimensional finite-element representations. Euler-Bernoulli beam formulations have been employed for nonlinear aeroelastic analysis of highly flexible aircraft, where the wing and tail structures are represented using geometrically nonlinear beam equations \citep{SHAFAGHAT2022107663}. Similarly, \citep{VINDIGNI2025107618} have proposed an aeroelastic beam finite element based on \textit{Euler-Bernoulli} and \textit{De-Saint-Venant} theories for flexible lifting structures. More general beam representations based on \textit{Timoshenko} theory have been incorporated into fully coupled aeroelastic solvers, including finite-element formulations for elastic wings and low-fidelity aeroelastic optimization frameworks \citep{KAMPCHEN200363,GILLEBAART2016512}. For plate-like structures, classical \textit{Kirchhoff–Love} formulations \citep{RAGAB2019299} provide an efficient representation when transverse shear effects are negligible. At higher fidelity, three-dimensional finite-element descriptions can represent complex structural details and spatially varying properties, although their computational cost may become restrictive during iterative design. Conversely, reduced-order and modal structural representations, provide substantial computational savings at the expense of restricting the structural response to a prescribed set of modes or reduced degrees of freedom. \citep{LI2024104055} propose a reduced order model for aerodynamic distribution prediction, used for static aeroelastic analysis with geometric nonlinearity. \textit{Reissner-Mindlin} plate formulations provide an intermediate description by incorporating transverse shear deformation while also retaining the computational advantages of a 2-D structural discretization \citep{PEREIRA2025113185,RAGAB2019299}. Their compatibility with standard  $C^0$ finite-element interpolation also makes them attractive for aeroelastic analysis involving general geometries.

For partitioned aeroelastic simulations, the accuracy of the interface transfer is not determined solely by the pointwise accuracy of interpolation, but the consistency between the exchanged loads and structural motion is also essential for preserving the energetics of the coupled system \citep{DEBOER20084284}. In particular, when the fluid and structural discretizations are non-conforming, an interface transfer that does not preserve the work performed by the aerodynamic loads can introduce artificial energy or remove energy from the coupled system, potentially affecting the stability and accuracy of the transient response. Conservative load and motion transfer methods for non-conforming interfaces have therefore received considerable attention in computational aeroelasticity \citep{farhat_workcons}, with formulations developed to preserve momentum and energy while allowing independent fluid and structural discretizations. Common-refinement approaches \cite{crm_initial} provide a geometric construction for transferring interface loads between non-matching meshes while satisfying discrete conservation properties, whereas radial basis function-based methods \citep{LOMBARDI2013117} provide a flexible mesh-independent interpolation framework that has also been adopted in aeroelastic and fluid-structure interaction simulations \cite{LIU2017810,aeroelasticmdo}. In this proposal, the work conservative interface treatment combines a parameterized representation of the communication surface with interpolation between non-matching parametric discretizations, between \textit{rVPM}-based aerodynamics and finite element-based structural discretization. Our framework is designed to accommodate independent non-matching interfaces while enabling the conservation and transfer accuracy of the two interface treatments to be assessed within the same aeroelastic formulation. Our primary contributions are as follows:

\begin{itemize}
    \item We develop a modular, open-source variable fidelity, aeroelastic framework that couples the numerically stable reformulated vortex particle method (\textit{rVPM}) using \texttt{FLOWUnsteady}, with a finite-element structural solver based on \texttt{FEniCS}, providing a partitioned aeroelastic framework.
    
    \item We demonstrate the application of \textit{rVPM} to aeroelastic dynamics, addressing the numerical-stability challenges that have historically limited the broader adoption of vortex-particle methods in coupled aeroelastic simulations. To the authors’ knowledge, this represents the first implementation of the \textit{rVPM} within an aeroelastic solver, with extensive validation of the coupled framework, including aerodynamic validation against water-tunnel measurements for a canonical heaving configuration and verification of the accuracy, convergence, and work conservative properties of the fluid–structure transfer.
    \item We extend the \texttt{FEniCS} framework along with the \texttt{fenics\_shells} library, to fully dynamic aeroelastic coupling through a Generalized-$\alpha$ time-integration scheme and a residual-based formulation that accounts explicitly for structural inertia, damping, and stiffness, in contrast to quasi-static coupling approaches such as S. P. van Schie et al. \cite{aeroelasticmdo}, based primarily on time-step-wise structural equilibrium and data exchange.

    \item We investigate non-matching interface transfer using both compactly supported radial-basis-function (RBF) interpolation and a common-refinement operator approach, providing a consistent framework for exchanging loads and displacements between aerodynamic control panels and structural finite elements. 

\end{itemize}

The remainder of this paper is structured as follows. Section. (\ref{sec:background}) establishes the theoretical foundation, presenting the constituent formulations for both the aerodynamic and structural models, followed by the time-integration scheme employed for the dynamic aeroelastic system. Section. (\ref{sec:comp}) outlines the computational framework, detailing the algorithmic workflow, spatial and temporal discretization, geometric parametrization, and boundary conditions. In the same section, we address the critical aspect of energy conservation at the fluid-structure interface, introducing and comparing two distinct work-conserving methodologies. Finally, Section. (\ref{sec:expt}) validates the proposed numerical solver through comprehensive numerical experiments, including validation cases, parametric sweeps, and convergence analyses. Section. (\ref{sec:conc}) concludes the paper with a summary of key findings and outlines directions for future research.

\section{Background Work}
\label{sec:background}

\subsection{Aerodynamic Model}
\label{sec:aero_model}

The aerodynamic solver in this work is built on the \textit{reformulated Vortex Particle Method} (\textit{rVPM}) \citep{alvarez2022flowunsteady}. This formulation models the fluid domain by solving the LES-filtered vorticity transport equation in a Lagrangian framework \citep{vortex_methods}. Continuous vorticity is discretized onto particles that simultaneously advect and stretch, eliminating the need for a volumetric computational grid. Consequently, the fluid solver circumvents dynamic remeshing overhead during structural deformations. Furthermore, Lagrangian transport preserves trailing wake vortical structures over long distances with minimal numerical dissipation \citep{vortex_winckelmans}.

In a meshless vortex particle solver, solid boundaries cannot be enforced via body-fitted grids. To introduce the structural body into the particle domain, an Actuator Surface Model (ASM) is employed within the \texttt{FLOWUnsteady} framework \citep{ASM, alvarez2022flowunsteady}. The ASM extends classical Vortex Lattice Method (VLM) principles \citep{Katz_Plotkin_2001} by representing the wing camber surface as a 2D lattice. Bound vorticity is shed from the trailing edge into the wake as free vortex particles at each time step. The induced velocity field and bound circulation are then used to evaluate the surface aerodynamic loads. The governing continuum equations, numerical particle dynamics, and force evaluations are detailed in the following subsections.

\begin{table}[pos=h!]
\centering
\caption{Notation and conventions for the fluid solver.}

\label{tab:fluid_notation}
\renewcommand{\arraystretch}{1.25}
\begin{tabular}{|c|p{9.5cm}|}

\hline
\textbf{Symbol} & \textbf{Description} \\
\hline

$\bm{\omega}(\mathbf{x})$ & Vorticity field \\

$\mathbf{u}(\mathbf{x})$ & Velocity field \\

$\overline{a}(\mathbf{x})$ & Filtered field $a(\mathbf{x})$ \\

$\bm{\Gamma}_{p}$ & Circulation vector of vortex particle $p$ \\

$\mathbf{x}_{p}$ & Position vector of vortex particle $p$ \\

$\sigma_{p}$ & Core radius (smoothing length) of vortex particle $p$ \\

$\hat{\mathbf{p}}$ & Unit vector along the vortex tube axis \\

$\nu_f$ & Kinematic viscosity of the fluid \\

$\rho_f$ & Fluid density \\

$\zeta_{\sigma}$ & Radial basis smoothing kernel of width $\sigma$ \\

$\mathbf{u}_{\infty}$ & Freestream velocity vector \\

$\mathbf{u}_{\mathrm{kin}}$ & Kinematic velocity induced by rigid body motion\\

$\mathbf{u}_{\mathrm{bound}}, \mathbf{u}_{\mathrm{wake}}$ & Velocity induced by bound vortex elements and free wake particles \\

$\mathbf{E}_{\mathrm{adv}}, \mathbf{E}_{\mathrm{str}}$ & Modeled Sub-Filter Scale Tensors for advection and stretching \\

$\mathbf{F}_{\mathrm{KJ}}, \mathbf{F}_{\mathrm{unsteady}}, \mathbf{F}_{\mathrm{parasitic}}$ & Kutta-Joukowski (circulatory), unsteady, parasitic force components \\ 

$f,\,g$ & Parameters controlling angular‑momentum / mass conservation \\

$\eta^s$ & Non‑dimensional spanwise coordinate $\in [0,1]$ \\

\hline
\end{tabular}
\end{table}

\subsubsection{Vorticity formulation and LES filtering}
\label{sec:vorticity_formulation}

Our aeroelastic framework, \texttt{VarFlExI}, models low-Mach number incompressible viscous flows. Taking the curl of the incompressible momentum Navier-Stokes equations yields the governing vorticity transport equation:

\begin{equation}
  \frac{\partial \bm{\omega}}{\partial t}
  + (\mathbf{u}\cdot\nabla) \bm{\omega}
  = (\bm{\omega}\cdot\nabla)\mathbf{u}
  + \nu_f \nabla^{2}\bm{\omega},
  \label{eq:vorticity_form}
\end{equation}

where $\bm{\omega} = \nabla \times \mathbf{u}$ is the vorticity vector field. Applying the curl operation eliminates the pressure gradient term, as the curl of any scalar gradient is identically zero ($\nabla \times \nabla p = \mathbf{0}$). Furthermore, the baroclinic torque term vanishes under the assumption of uniform fluid density ($\nabla \rho_f = \mathbf{0}$). Equation \eqref{eq:vorticity_form} establishes that vorticity changes in time due to convective transport $(\mathbf{u}\cdot\nabla)\bm{\omega}$, vortex stretching and tilting $(\bm{\omega}\cdot\nabla)\mathbf{u}$, and viscous diffusion $\nu_f \nabla^{2}\bm{\omega}$.

Direct numerical simulation of high-Reynolds-number turbulent wakes is computationally prohibitive for unsteady flow. Therefore, a Large-Eddy Simulation (LES) approach is used to resolve large-scale coherent structures while accounting for unresolved small-scale dynamics \citep{mansfield_vortex}. A low-pass spatial filter is applied to the flow domain. The filtered vorticity field $\overline{\bm{\omega}}(\mathbf{x},t)$ is defined mathematically by the spatial convolution:

\begin{equation}
  \overline{\bm{\omega}}(\mathbf{x},t)
  \equiv \int_{\mathbb{R}^{3}} \bm{\omega}(\mathbf{y},t)\,
  \zeta_{\sigma}(\mathbf{x}-\mathbf{y})\,\mathrm{d}\mathbf{y},
  \label{eq:les_filter}
\end{equation}

where $\zeta_{\sigma}$ represents a localized, radially symmetric smoothing kernel of characteristic filter width $\sigma$. The filter parameter $\sigma$ defines the spatial resolution scale separating resolved flow motions from subfilter scales (SFS) \citep{winckelmans}. Applying the filtering operator \eqref{eq:les_filter} to the transport equation \eqref{eq:vorticity_form} introduces non-linear commutator errors. In general, the filter of a product of field quantities does not equal the product of their filtered components. Consequently, the non-linear advection ($u_j \frac{\partial \omega_i}{\partial x_j}$) and vortex stretching ($\omega_j \frac{\partial u_i}{\partial x_j}$) terms cannot be evaluated directly from resolved quantities. Instead, they are modeled through a residual subfilter stress tensor $T_{ij}$ \citep{mansfield_vortex, alvarez2022flowunsteady}. This tensor encapsulates the subfilter interactions between the filtered product and the product of resolved fields,

\begin{equation}
  T_{ij} = \overline{u_i \omega_j} - \overline{u}_i \overline{\omega}_j.
  \label{eq:sfs_tensor_def}
\end{equation}

Taking the spatial gradients of $T_{ij}$ and its transpose $T'_{ij} = T_{ji}$ isolates the unresolved advective and stretching contributions,

\begin{align}
  \frac{\partial T'_{ij}}{\partial x_j} 
  &= \overline{u_j \frac{\partial \omega_i}{\partial x_j}} 
   - \overline{u}_j \frac{\partial \overline{\omega}_i}{\partial x_j}, \label{eq:sfs_adv_grad} \\[1ex]
  \frac{\partial T_{ij}}{\partial x_j} 
  &= \overline{\omega_j \frac{\partial u_i}{\partial x_j}} 
   - \overline{\omega}_j \frac{\partial \overline{u}_i}{\partial x_j} \label{eq:sfs_str_grad}
\end{align}

Substituting Eqs.~\eqref{eq:sfs_adv_grad} - \eqref{eq:sfs_str_grad} into the filtered vorticity equation yields the LES vorticity transport equation in index notation:

\begin{equation}
  \frac{\partial \overline{\omega}_i}{\partial t} 
  + \overline{u}_j \frac{\partial \overline{\omega}_i}{\partial x_j} 
  = \overline{\omega}_j \frac{\partial \overline{u}_i}{\partial x_j} 
  + \nu_f \nabla^2 \overline{\omega}_i 
  - \frac{\partial T'_{ij}}{\partial x_j} 
  + \frac{\partial T_{ij}}{\partial x_j}
  \label{eq:les_vorticity_index}
\end{equation}

The term $\frac{\partial T'_{ij}}{\partial x_j}$ represents the subfilter-scale advective flux ($\mathbf{E}_{\mathrm{adv}}$), while $\frac{\partial T_{ij}}{\partial x_j}$ represents the subfilter vortex stretching flux ($\mathbf{E}_{\mathrm{str}}$). These terms capture the enstrophy transfer between the resolved and unresolved scales. In the context of vorticity dynamics, however, treating subfilter effects as purely diffusive fails to capture the directional energy cascade \citep{FLOWUnsteady}. 

In the present simulations, the unresolved-scale effects are represented by the implicit regularization associated with the finite-width particle kernel and its dynamically evolving core size $\sigma_p$.

\subsubsection{Lagrangian Discretisation}

The continuous LES spatial filtering integral in Eq.~\eqref{eq:les_filter} is discretized by approximating the vorticity field as a finite ensemble of $N$ Lagrangian particles \citep{vortex_methods, leonard_vortex}. Each particle has a circulation $\bm{\Gamma}_p(t)$ and a localized core radius $\sigma_p(t)$. Evaluating the filtering convolution \eqref{eq:les_filter} via particle quadrature yields the discrete filtered vorticity field:

\begin{equation}
  \overline{\bm{\omega}}(\mathbf{x},t)
  \approx \sum_{p}
  \bm{\Gamma}_{p}(t)\,
  \zeta_{\sigma_{p}}\bigl(\mathbf{x}-\mathbf{x}_{p}(t)\bigr),
  \label{eq:particle_discrete}
\end{equation}

where $\mathbf{x}_p(t)$ denotes the position vector of particle $p$. Through this quadrature formulation, the particle core radius $\sigma_p$ directly represents the localized spatial filter width $\sigma$ introduced in Eq.~\eqref{eq:les_filter}.

Vortex particles move along fluid trajectories with the local velocity $\mathbf{u}(\mathbf{x}_p)$. Unlike classical fixed-core particle formulations,
the \textit{rVPM} allows both the circulation vector $\bm{\Gamma}_p$ and core radius $\sigma_p$ to evolve dynamically in response to local vortex stretching and viscous diffusion \citep{alvarez2022flowunsteady}. Because $\sigma_p$ acts as the local filter width, its evolution modulates the local spatial resolution through particle dilation and contraction. In the present
simulations, the finite-width and dynamically evolving particle kernel provides the implicit regularization of the resolved vorticity field.

\subsubsection{Governing Equations of the \textit{rVPM}}
\label{sec:rVPM_governing}

Substituting the discrete particle summation \eqref{eq:particle_discrete} into the filtered transport equation \eqref{eq:les_vorticity_index} yields the governing ordinary differential equations (ODEs) of the \textit{rVPM} \citep{alvarez2022flowunsteady}. The dynamic state of each vortex particle $p$ is defined by three primary variables: its position vector $\mathbf{x}_p$, its core radius $\sigma_p$, and its circulation vector $\bm{\Gamma}_p$. Selecting the dimensionless parameters $(f,g)$ determines the balance in exact conservation of angular momentum against conservation of vortex-tube volume (mass), and hence the \textit{rVPM} transport model:

\begin{align}
  \frac{\mathrm{d}\mathbf{x}_{p}}{\mathrm{d}t}
  &= \mathbf{u}(\mathbf{x}_{p}), \label{eq:rVPM_x} \\[2ex]
  \frac{\mathrm{d}\sigma_{p}}{\mathrm{d}t}
  &= -\,\frac{g+f}{1+3f}\,
     \frac{\sigma_{p}}{\|\bm{\Gamma}_{p}\|}
     \bigl[(\bm{\Gamma}_{p}\!\cdot\!\nabla)\mathbf{u}(\mathbf{x}_{p})\bigr]
     \!\cdot\!\hat{\mathbf{p}}, \label{eq:rVPM_sigma} \\[2ex]
  \frac{\mathrm{d}\bm{\Gamma}_{p}}{\mathrm{d}t}
  &= (\bm{\Gamma}_{p}\!\cdot\!\nabla)\mathbf{u}(\mathbf{x}_{p})
     - \frac{g+f}{\tfrac{1}{3}+f}
     \Bigl\{\bigl[(\bm{\Gamma}_{p}\!\cdot\!\nabla)\mathbf{u}(\mathbf{x}_{p})\bigr]
     \!\cdot\!\hat{\mathbf{p}}\Bigr\}\,\hat{\mathbf{p}},
  \label{eq:rVPM_Gamma}
\end{align}

where $\hat{\mathbf{p}} = \frac{\bm{\Gamma}_{p}}{\|\bm{\Gamma}_{p}\|}$ is the unit vector oriented along the local vortex tube axis. The velocity $\mathbf{u}(\mathbf{x}_{p})$ represents the total velocity field induced at the particle location. Each evolution equation models a distinct physical mechanism:

\begin{enumerate}
  \item Equation \eqref{eq:rVPM_x}: Governs the spatial trajectory of particle $p$. Particles advect passively along fluid streamlines with the local velocity vector $\mathbf{u}(\mathbf{x}_p)$.
  
  \item Equation \eqref{eq:rVPM_sigma}: Controls particle core deformation due to inviscid vortex stretching. The term $\bigl[(\bm{\Gamma}_p \!\cdot\!\nabla)\mathbf{u}(\mathbf{x}_p)\bigr] \!\cdot\!\hat{\mathbf{p}}$ measures the directional velocity gradient (stretching or compression rate) along the vortex axis $\hat{\mathbf{p}}$. Positive axial stretching forces the core radius $\sigma_p$ to contract, whereas axial compression causes the core radius to dilate.
  
  \item Equation \eqref{eq:rVPM_Gamma}: Governs the magnitude and directional reorientation of the particle circulation vector. The first term $(\bm{\Gamma}_p \!\cdot\!\nabla)\mathbf{u}(\mathbf{x}_p)$ represents classical 3D vortex stretching and tilting. The second term subtracts a scaled projection along $\hat{\mathbf{p}}$ to balance the strain-induced core radius variation in Eq.~\eqref{eq:rVPM_sigma}.
\end{enumerate}

In the present work, parameters are set to $f=0$ and $g=1/5$. This selection enforces a momentum-conserving sphere during vortex stretching operations \citep{FLOWUnsteady}. The coupled ODE system Eqs. \eqref{eq:rVPM_x}-\eqref{eq:rVPM_Gamma} dictates how particle properties evolve over time in turbulent flow fields. When a vortex tube experiences positive axial strain, the circulation magnitude $\|\bm{\Gamma}_p\|$ increases while the core radius $\sigma_p$ simultaneously contracts. Because $\sigma_p$ defines the local spatial filter width, core contraction sharpens local spatial resolution, driving enstrophy transfer toward smaller, higher wavenumber resolved scales (forward cascade). Conversely, when a particle encounters axial compression, $\|\bm{\Gamma}_p\|$ decreases while $\sigma_p$ dilates. This local filter expansion merges small-scale interactions back into broader flow structures (inverse cascade/backscatter).

The \textit{rVPM} treatment of unresolved-scale dynamics relies on the regularization associated with the finite-width particle kernel and the evolving particle core, while viscous diffusion is incorporated through
particle core spreading as given in Appendix \ref{app:viscous}.








\subsubsection{Lifting Surface and Wake Coupling}
\label{sec:rVPM_coupling}

In a meshless Lagrangian solver, solid boundaries cannot be represented by body-fitted volume grids. Therefore, the total vorticity field $\bm{\omega}$ is decomposed into a bound component tied to the wing geometry and a free component representing the wake:

\begin{equation}
  \bm{\omega}(\mathbf{x},t) = \bm{\omega}_{\mathrm{bound}}(\mathbf{x},t) + \bm{\omega}_{\mathrm{free}}(\mathbf{x},t)
  \label{eq:vorticity_decomposition}
\end{equation}

To represent the deforming wing structure, an Actuator Surface Model (\textit{ASM}) is implemented within the \texttt{FLOWUnsteady} framework \citep{alvarez2022flowunsteady}. Standard Vortex Lattice Methods (\textit{VLM}) represent bound vorticity as infinitely thin singular filaments \citep{Katz_Plotkin_2001}. However, singular filaments induce unbounded velocity spikes when free wake particles advect near the surface. The \textit{ASM} resolves this numerical instability by regularizing the bound circulation across the surface using a spatial kernel $\zeta_{\sigma}$ \citep{ASM, alvarez2022flowunsteady}.

Bound vortex elements are positioned along the quarter-chord line ($\eta_{\mathrm{bv}} = 1/4$) of each local element. Kinematic no-penetration boundary conditions are enforced at the control point located at the three-quarter-chord station ($\eta_{\mathrm{cp}} = 3/4$). The following locations for the points of interest are chosen from an empirical law for airfoil structures. According to \citep{pistolesi}, enforcing zero normal flow at the $3/4$-chord station satisfies the Kutta condition at the trailing edge for thin lifting surfaces. Solving the resulting linear system at each time step yields the bound circulation distribution $\Gamma_i$ for $i = 1, \dots, N_{\mathrm{span}}$. Unsteady circulation variations are subsequently shed from the trailing edge into the fluid domain as free vortex particles that advect according to Eqs.~\eqref{eq:rVPM_x}-\eqref{eq:rVPM_Gamma}. The total flow velocity $\mathbf{u}(\mathbf{x},t)$ at any spatial location is assembled via kinematic superposition:

\begin{equation}
  \mathbf{u} = \mathbf{u}_{\infty}
             + \mathbf{u}_{\mathrm{kin}}
             + \mathbf{u}_{\mathrm{bound}}
             + \mathbf{u}_{\mathrm{wake}},
  \label{eq:total_velocity}
\end{equation}

where $\mathbf{u}_{\infty}$ is the freestream velocity, $\mathbf{u}_{\mathrm{kin}}$ is the velocity induced by structural motion and elastic deformation, $\mathbf{u}_{\mathrm{bound}}$ is the velocity induced by the bound surface elements, and $\mathbf{u}_{\mathrm{wake}}$ is the velocity induced by free wake particles. Direct calculation of pairwise Biot-Savart interactions between $N$ wake particles and surface elements requires $\mathcal{O}(N^2)$ operations. To make long-duration unsteady aeroelastic simulations computationally tractable, mutual particle-particle and particle-surface velocity evaluations are accelerated using a Fast Multipole Method (FMM), reducing algorithm complexity to $\mathcal{O}(NlogN)$ \citep{fmm, AndrewFLOWFMM}.

\subsubsection{Aerodynamic Force Evaluation}
\label{sec:rVPM_forces}

The aerodynamic force acting on each bound surface element $i$ is calculated using the effective local velocity $\mathbf{u}_{\mathrm{local},i}$ evaluated at
its control point $\mathbf{x}_{\mathrm{cp},i}$,

\begin{equation}
  \mathbf{u}_{\mathrm{local},i} = \mathbf{u}(\mathbf{x}_{\mathrm{cp},i}) - \mathbf{u}_{\mathrm{self},i},
  \label{eq:ulocal}
\end{equation}

where $\mathbf{u}(\mathbf{x}_{\mathrm{cp},i})$ is the total velocity vector assembled via Eq.~\eqref{eq:total_velocity}, and $\mathbf{u}_{\mathrm{self},i}$ is the velocity self-induced by element $i$. This isolates the external cross-flow acting on the element, avoiding non-physical self-induced force singularities. The total aerodynamic force vector $\mathbf{F}_i$ on element $i$ is decomposed into circulatory, unsteady, and parasitic components:

\begin{equation}
  \mathbf{F}_i = \mathbf{F}_{\mathrm{KJ},i} + \mathbf{F}_{\mathrm{unsteady},i} + \mathbf{F}_{\mathrm{parasitic},i}
  \label{eq:force_decomposition}
\end{equation}

The quasi-steady circulatory load $\mathbf{F}_{\mathrm{KJ},i}$, which contributes primarily to $\mathbf{F}_i$, is evaluated using the generalized Kutta-Joukowski theorem \citep{Katz_Plotkin_2001},

\begin{equation}
  \mathbf{F}_{\mathrm{KJ},i} = \rho_f\,\Gamma_i \bigl(\mathbf{u}_{\mathrm{local},i} \times \mathbf{l}_i\bigr),
  \label{eq:kj_force}
\end{equation}

where $\rho_f$ is the fluid density, $\Gamma_i$ is the bound circulation, and $\mathbf{l}_i$ is the bound segment vector. Unsteady loads $\mathbf{F}_{\mathrm{uns},i}$ account for apparent mass effects, while parasitic drag $\mathbf{F}_{\mathrm{parasite},i}$ incorporates viscous skin friction and pressure drag from 2D sectional polars \citep{FLOWUnsteady}.

\begin{figure}[pos=h!]
    \centering
    \includegraphics[width=1.0\linewidth]{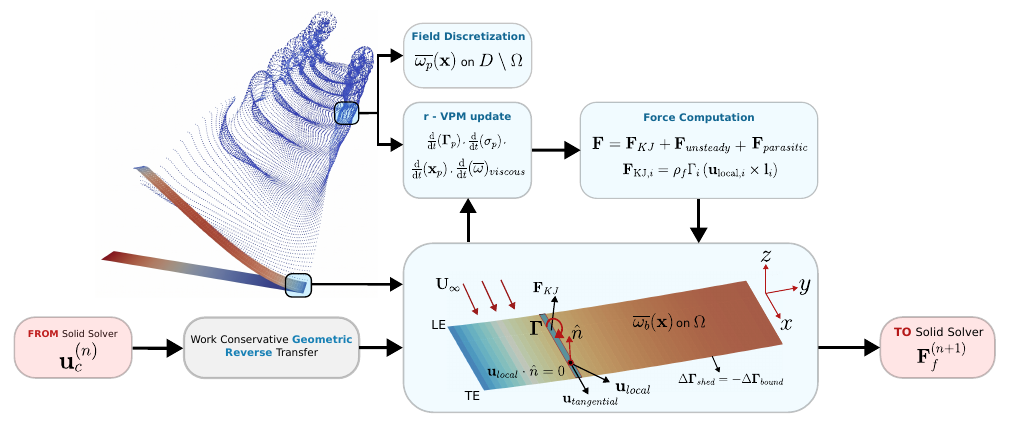}
    \caption{Schematic of the aerodynamic solver constructed using the open-source
    \texttt{FLOWUnsteady} library in \texttt{Julia}. The aerodynamic domain is represented using a regularized vortex-particle formulation, with the vorticity field discretized into vortex particles and evolved through the \textit{r-VPM} update.
    For the lifting-surface representation, each wing element contains a bound vortex carrying circulation $\Gamma$ at the quarter-chord, and a control
    point is located at the three-quarter of chord. The local velocity $\mathbf{u}_{\mathrm{local}}$ is evaluated at the control point through the decomposition given in Eq. \eqref{eq:total_velocity}, and the no-flow-through boundary condition is imposed there to determine the bound circulation distribution. The resulting aerodynamic loads are evaluated from the bound circulation (KJ), together with unsteady and parasitic contributions. Circulation shed from the trailing edge is represented through the corresponding change in bound and shed circulation. The figure also illustrates the bidirectional coupling with the structural solver, where structural displacements are transferred to the aerodynamic representation through the work-conservative geometric reverse transfer, while the computed aerodynamic loads are transferred back to the structural interface. The shedding visualization of the fluid panels shown represents aeroelastic flutter induced by the incoming freestream, simulated for a flexible structure ($E \approx 10^8$).}
    \label{fig:fluid_flowchart}
\end{figure}

To prevent unbounded growth of wake particles over time, the circulation magnitude and the core radii of particles are lower and upper bounded, and hence are pruned from the domain \citep{alvarez2022flowunsteady}. Because velocity induction decays as $1/r^2$ via the Biot-Savart law, pruning highly diffused downstream particles maintains fast FMM evaluations and prevents far-field numerical blockage without degrading the precision of $\mathbf{u}_{\mathrm{local},i}$ at the wing surface. In the upcoming section, we discuss the background work for the structural solver. Readers can refer to Fig. \ref{fig:fluid_flowchart} for a better understanding of the fluid solver architecture.

\subsection{Structural Model}
\label{sec:solid_model}

\begin{table}[pos=h!]
\centering
\caption{Notation and conventions - Solid Structural Solver}
\label{tab:nomenclature}
\renewcommand{\arraystretch}{1.25}
\begin{tabular}{|c|p{9.5cm}|}
\hline
\textbf{Symbol} & \textbf{Description} \\
\hline

$\mathbf{q}$ & Structural State vector, $
\mathbf q=
\left[
u_0,
v_0,
w,
\theta_x,
\theta_y
\right]^T
$\\

$w$ & Transverse Displacement \\

$\mathbf{u}$ & Nodal displacement vector field, $\mathbf{u}=[u,v]^T$\\

$\mathbf{\theta}$ & Independent Rotational DoF's, $\mathbf{\theta}=[\theta_x,\theta_y]^T$\\

$\rho_s$ & Structural density \\

$\boldsymbol{\varepsilon},\kappa$ & Membrane strain \& Bending Curvature Tensor \\

$E$ & Young’s modulus \\

$\nu$ & Poisson’s ratio \\

$\lambda,\mu$ & Lamé elastic constants \\

$\mathcal{M},\,\mathcal{C},\,\mathcal{K},\,\mathcal{W}_{\mathrm{ext}}$
& Inertial, damping, internal elastic, and external virtual-work functionals. \\

$\mathbf{F_S}$ & External nodal force vector \\

$\eta_m,\eta_k$ & Rayleigh damping coefficients \\

$\alpha_m, \alpha_f$ & Generalized-$\alpha$: inertia and force weighting parameters \\

$\Gamma_s$ & Structural domain \\

\hline
\end{tabular}
\end{table}

Modeling the structural response using the full three-dimensional theory of elasticity requires solving the displacement field throughout the plate thickness, resulting in three displacement unknowns at every point in the domain. Although highly accurate, this approach becomes computationally prohibitive due to the large number of elements required through the thickness. Classical \textit{Kirchhoff-Love} plate theory reduces this computational cost by discretizing on a two-dimensional mid-surface \citep{3dviscous_nonlinearpanel}; however, it assumes that normals to the mid-surface remain straight and perpendicular after deformation, thereby neglecting transverse shear deformation \citep{nguyen2021stable,GORDNIER2002497}. While this assumption is appropriate for very thin plates, it suppresses transverse shear deformation and therefore becomes increasingly inaccurate for moderately thick structures, for which transverse shear contributes appreciably to the overall deformation and compliance, potentially resulting in an artificially stiff structural response \citep{article_reisner_kirchoff}. 

To accurately capture these effects while retaining the computational efficiency of a two-dimensional formulation, our work employs the \textit{Reissner-Mindlin} plate theory, which introduces independent rotational DoFs of the plate normals, allowing transverse shear deformation to be represented without resorting to a full three-dimensional elasticity formulation \citep{article_reisner}. Compared to classical \textit{Kirchhoff-Love} plate theory, the \textit{Reissner-Mindlin} formulation retains transverse shear effects \citep{article_reisner_kirchoff}, making it suitable for moderately thick wings. The formulation describes the structural response through the mid-surface displacement and rotation fields, thereby incorporating membrane, bending, and transverse shear deformation while reducing the three-dimensional elasticity problem to a two-dimensional representation of the plate. The following sections describe the modeling assumptions made for the structural solver, the kinematic model description, and the time integration method.

\subsubsection{Kinematic Modeling Assumptions}
The plate is assumed to be homogeneous, isotropic, and of uniform thickness
$h$, and initially planar \citep{influence_rotatory,formulation_thickthin}. The plate thickness is assumed to be small relative to its characteristic in-plane dimensions. The middle surface of the plate occupies the domain
$\Omega \subset \mathbb{R}^2$, while the thickness coordinate is denoted
by $z \in [-h/2,h/2]$. Under these assumptions, the displacement field is decomposed into in-plane membrane displacements and out-of-plane transverse deflection, in terms of the
displacement and independent rotations of the middle surface. The 3-dimensional displacement field is therefore approximated as

\begin{equation}
u(x,y,z)=u_0(x,y)-z\theta_x(x,y),
\quad
v(x,y,z)=v_0(x,y)-z\theta_y(x,y),
\quad 
w(x,y,z)=w(x,y),
\end{equation}

where $[v_0,u_0,w]^T$ is the mid-surface displacement and $[\theta_x,\theta_y]^T$ are the independent rotations. According to the \textit{Reissner-Mindlin} kinematic assumptions, points initially lying on a normal to the mid-surface remain on a straight line after deformation, but the normal is not constrained to remain perpendicular to the deformed mid-surface. Consequently, independent rotations are introduced to account for transverse shear deformation in the unknown structural , $
\mathbf q=
\left[
u_0,
v_0,
w,
\theta_x,
\theta_y
\right]^T
$. The deformation is decomposed into membrane, bending, and transverse shear components. The membrane strain tensor is

\begin{equation}
\boldsymbol{\varepsilon}
=
\frac12
\left(
\nabla \mathbf u_0
+
\nabla \mathbf u_0^T
\right); \quad
\mathbf u_0=
\begin{bmatrix}
u_0\\
v_0
\end{bmatrix}
\end{equation}

The bending curvature tensor is obtained from the spatial variation of the plate rotations,

\begin{equation}
\boldsymbol{\kappa}
=
\frac12
\left(
\nabla\boldsymbol{\theta}
+
\nabla\boldsymbol{\theta}^{T}
\right);\quad
\boldsymbol{\theta}
=
\begin{bmatrix}
\theta_x\\
\theta_y
\end{bmatrix}
\end{equation}

Unlike \textit{Kirchhoff-Love} plate theory, transverse shear deformation is retained through $
\boldsymbol{\gamma}
=
\nabla w
-
\boldsymbol{\theta},
$ which represents the relative rotation between the plate normal and the gradient of the transverse displacement. The membrane stress $\mathbf{N}$, bending moment tensor $\mathbf{B}$ and transverse force resultant $\mathbf{Q}$ are

\begin{equation}
\mathbf N
=
\frac{Eh}{1-\nu^2}
\left[
(1-\nu)\boldsymbol{\varepsilon}
+
\nu
\,
\mathrm{tr}(\boldsymbol{\varepsilon})
\mathbf I
\right],\quad 
\mathbf B
=
\frac{Eh^3}
{12(1-\nu^2)}
\left[
(1-\nu)\boldsymbol{\kappa}
+
\nu
\,
\mathrm{tr}(\boldsymbol{\kappa})
\mathbf I
\right],\quad
\mathbf Q
=
\kappa_sGh
\,
\boldsymbol{\gamma},
\end{equation}

where $E$ denotes Young's modulus, $\nu$ is Poisson's ratio, $G
=\frac{E}{2(1+\nu)}$ is the Shear modulus, and $\kappa_s$ is the shear correction factor. Given the state $
\mathbf q=
\left[
u_0,
v_0,
w,
\theta_x,
\theta_y
\right]^T
$, the governing equations are obtained from the principle of virtual work, for an arbitrary virtual displacement $\delta\mathbf q$, and the weak form can be written as

\begin{equation}
\mathcal{M}(\ddot{\mathbf q},\delta\mathbf q)
+
\mathcal{C}(\dot{\mathbf q},\delta\mathbf q)
+
\mathcal{K}(\mathbf q,\delta\mathbf q)
=
\mathcal{W}_{\mathrm{ext}}(\delta\mathbf q),
\end{equation}

where $\mathcal{M}$, $\mathcal{C}$, $\mathcal{K}$ and
$\mathcal{W}_{\mathrm{ext}}$ denote the inertial, damping, and internal and external virtual-work functionals, respectively. The inertial contribution is given by

\begin{equation}
\mathcal M
=
\int_\Omega
\rho_s h
\left(
\mathbf u\cdot\delta\mathbf u
+
w\delta w
\right)
d\Omega
+
\int_\Omega
\frac{\rho_s h^3}{12}
\,
\boldsymbol{\theta}
\cdot
\delta\boldsymbol{\theta}
\,
d\Omega
\end{equation}

The elastic contribution consists of membrane, bending, and transverse shear energies, $
\mathcal K
=
K_m
+
K_b
+
K_s
$, where

\begin{equation}
K_m
=
\int_\Omega
\mathbf N:
\delta\boldsymbol{\varepsilon}
\,d\Omega;\quad
K_b
=
\int_\Omega
\mathbf B:
\delta\boldsymbol{\kappa}
\,d\Omega;\quad
K_s
=
\int_\Omega
\mathbf{Q}
\cdot
\delta\boldsymbol{\gamma}
\,d\Omega
\end{equation}

To account for structural energy dissipation, a Rayleigh proportional damping method is adopted \citep{rayleigh}. The damping matrix is represented as a linear
combination of the mass and stiffness matrices, $
\mathcal C
=
\eta_m\mathcal M
+
\eta_k\mathcal K,
$ where $\eta_m$ and $\eta_k$ are the mass- and stiffness-proportional
damping coefficients, respectively. The mass-proportional component
primarily contributes to damping at lower frequencies, whereas the stiffness-proportional component produces increasing damping with
frequency. This representation preserves modal orthogonality when $\mathcal M$ and $\mathcal K$ are simultaneously diagonalizable, making it convenient for finite-element structural dynamics. The coefficients $\eta_m$ and $\eta_k$ may be selected to reproduce prescribed damping ratios over the frequency range of interest. For the $i$-th vibration mode with natural frequency $\omega_i$, the corresponding modal damping ratio is

\begin{equation}
\zeta_i
=
\frac{1}{2}
\left(
\frac{\eta_m}{\omega_i}
+
\eta_k\omega_i
\right)
\end{equation}

Upon spatial discretization using finite elements, the corresponding
virtual-work contributions give rise to the mass, damping, and
structural stiffness matrices. Readers can refer to Appendix (\ref{app:work_to_nodal_force}) for the equivalence of these notations.

\subsubsection{Time Integration }
The transient structural response is integrated using the \textit{Generalized-$\alpha$ method} \citep{gen-alpha,structural_dynamics}. The method provides second-order accuracy and, for an appropriate choice of its algorithmic parameters, unconditional stability for linear structural systems. In addition, it introduces controllable numerical dissipation that preferentially damps spurious high-frequency components while preserving the response of the low-frequency modes that dominate the physical structural dynamics. These properties make it particularly suitable for long-duration transient simulations and partitioned FSI problems \citep{gen_alpha,li2019novel}, where numerical stability is essential while preserving the physically relevant structural dynamics. The effectiveness of the adopted time integration scheme is further demonstrated through the validation studies presented in Section. (\ref{sec:validation}). The semi-discrete governing equation of motion is given as:

\begin{equation}
\mathbf M \ddot{\mathbf{q}} + \mathbf C \dot{\mathbf{q}} + \mathbf K \mathbf{q} = \mathbf F_s,
\label{eqn:final_solid_eqn}
\end{equation}
where $\mathbf{M}$ and $\mathbf{C}$ denote the mass and damping matrices, $\mathbf{F}_{s}$ represents the externally applied force vector, including the aerodynamic loads transferred from the fluid solver. The semi-discrete form of Eq.~\eqref{eqn:final_solid_eqn} follows from the finite element discretization of the weak form presented in the previous section. In particular, the external virtual work is transformed into an equivalent global nodal force vector through the finite element approximation of the displacement field. In the present partitioned framework, however, this nodal force vector is obtained directly from the conservative work force-transfer procedure as described in Section. (\ref{sec:work_transfer}). Consequently, the external loading appears explicitly as the force vector $\mathbf{F}_s$ in Eq.~\eqref{eqn:final_solid_eqn}, eliminating the need to reconstruct and integrate a continuous traction field over the structural surface. 
For a time interval $[0,T]$, the solution is advanced over $N$ uniform time steps of size $\Delta t=T/N$, with discrete time levels $t_0=0,t_1,\ldots,t_N=T$. Within the Generalized-$\alpha$ framework, the structural equilibrium is enforced at intermediate time levels between $t_n$ and $t_{n+1}$ according to

\begin{equation}
\label{gen_alpha}
\mathbf M\ddot{\mathbf q}_{n+1-\alpha_m}
+
\mathbf C\dot{\mathbf q}_{n+1-\alpha_f}
+
\mathbf K\mathbf q_{n+1-\alpha_f}
=
\mathbf F_{s,n+1-\alpha_f}
\end{equation}

Following the convention
adopted here, an intermediate quantity is defined as

\begin{equation}
\mathbf X_{n+1-\alpha}
=
(1-\alpha)\mathbf X_{n+1}
+
\alpha\mathbf X_n,
\end{equation}

with $\mathbf X\in{\mathbf q,\dot{\mathbf q},\ddot{\mathbf q},\mathbf F}$. The parameters $\alpha_f$ and $\alpha_m$ are associated with the force and inertia terms, respectively. We use '$\alpha_f$' for force/stiffness terms and '$\alpha_m$' for inertia terms. The displacement and velocity updates follow the Newmark-type approximations \citep{structural_dynamics}:
\begin{equation} 
    \mathbf{q}_{n+1} = \mathbf{q}_n + \Delta t \dot{\mathbf{q}}_n + \frac{\Delta t^2}{2} \left[ (1-2\beta)\ddot{\mathbf{q}}_n + 2\beta \ddot{\mathbf{q}}_{n+1} \right], \\ \label{newmark} \dot{\mathbf{q}}_{n+1} = \dot{\mathbf{q}}_n + \Delta t \left[ (1-\gamma)\ddot{\mathbf{q}}_n + \gamma \ddot{\mathbf{q}}_{n+1} \right]
\end{equation}

Rearranging the displacement update equation yields the acceleration at the new time level: 
\begin{equation} \ddot{\mathbf{q}}_{n+1} = \frac{1}{\beta \Delta t^2} \left( \mathbf{q}_{n+1} - \mathbf{q}_n - \Delta t \dot{\mathbf{q}}_n \right) - \frac{1-2\beta}{2\beta} \ddot{\mathbf{q}}_n
\label{eq:semi-discrete eqlm}
\end{equation}

The velocity is subsequently updated using Eq.~\eqref{newmark}. These updated kinematic quantities are evaluated at the Generalized-$\alpha$ intermediate states and substituted into Eq.~\eqref{gen_alpha} to form the residual of the semi-discrete equilibrium equations. The Generalized-$\alpha$ parameters are selected to achieve second-order accuracy and controlled high-frequency numerical dissipation. In the present work, the parameters are chosen as,

\begin{equation} \gamma = \frac{1}{2} + \alpha_f - \alpha_m, \qquad \beta = \frac{1}{4} \left( 1+\gamma-\alpha_f \right)^2,
\end{equation}

where $\alpha_m=0.1,\alpha_f=0.2$. Rather than forming an effective stiffness matrix, our implementation assembles the residual and its corresponding tangent Jacobian directly, following the finite element implementation described in the subsequent section. Substituting the Generalized-$\alpha$ intermediate displacement, velocity, and acceleration fields into Eq.~\eqref{gen_alpha} yields the residual form of the semi-discrete equilibrium equation,

\begin{equation}
\mathbf{R}_{n+1-\alpha_f}(\mathbf{q})
=
\mathbf{M}\ddot{\mathbf{q}}_{n+1-\alpha_m}
+
\mathbf{C}\dot{\mathbf{q}}_{n+1-\alpha_f}
+
\mathbf{K}\mathbf{q}_{n+1-\alpha_f}-
\mathbf{F}_{s,n+1-\alpha_f},
\label{eq:residual}
\end{equation}

where equilibrium is attained when $
\mathbf{R}(\mathbf{q})=\mathbf{0}
$. The nonlinear system is solved iteratively using a Gauss-Newton procedure. For the $k^{\mathrm{th}}$ iteration, the residual is linearized about the current estimate of the solution, yielding $
\mathbf{R}(\mathbf{q}^{k})
+
\mathbf{J}^{k}\Delta\mathbf{q}
=
\mathbf{0}
$, where $
\mathbf{J}^{k}
=
\frac{\partial\mathbf{R}}
{\partial\mathbf{q}}
\bigg|_{\mathbf{q}^{k}}
$ is the tangent Jacobian matrix evaluated at the current iterate. Solving the linearized system gives the displacement correction, $
\mathbf{J}^{k}\Delta\mathbf{q}
=
-\mathbf{R}(\mathbf{q}^{k}),
\label{eq:newton}
$ which is used to update the structural state according to $
\mathbf{q}^{k+1}
=
\mathbf{q}^{k}
+
\Delta\mathbf{q}$. The residual and Jacobian are reassembled using the updated structural state, and the Newton iterations are repeated until the prescribed convergence tolerances are satisfied. Upon convergence, the acceleration and velocity fields are updated using the Generalized-$\alpha$ relations before advancing to the next time step.

\subsubsection{Levenberg-Marquardt Damping}
To improve the robustness of the nonlinear solution procedure in
situations where the Newton iteration may exhibit poor convergence, a
\textit{Levenberg-Marquardt} (LM) damping strategy is considered. Rather than using the undamped Newton correction directly, the linearized residual equation is regularized by adding a positive diagonal term to the tangent system. At Newton iteration $k$, the correction $\Delta \mathbf{q}^{k}$ is obtained from $
    (J^T J+\lambda_{LM} I)\Delta q=-R
$ formulated through the least squares objective $\phi(q)=\frac{1}{2} || R(q)||^2$. This improves the robustness of the solutions when the current iterate is far from equilibrium or when the tangent Jacobian is poorly conditioned. The LM procedure is therefore used as a non-linear solution regularization within each Generalized-$\alpha$ timestep with $\Delta q=-(J^TJ+\lambda_{LM} I)^{-1}R$.

\begin{figure}[pos=h!]
    \centering
    \includegraphics[width=1.0\linewidth]{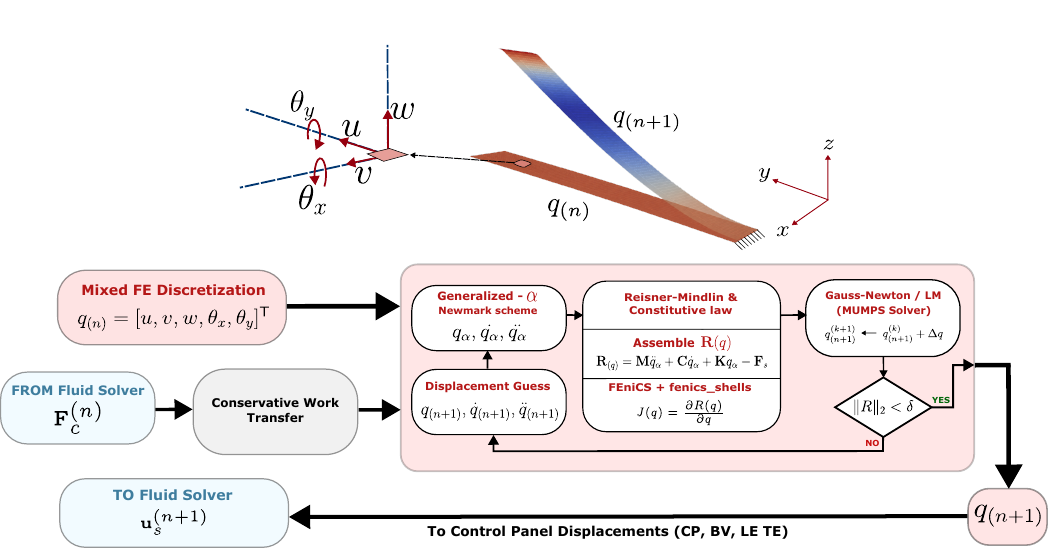}
    \caption{Schematic of the proposed partitioned structural solver. The
    \textit{\textbf{Reissner--Mindlin}} plate formulation in \texttt{FEniCS} is spatially discretized using
    finite elements and advanced in time with the \textit{\textbf{Generalized}}-$\alpha$
    method. At each time step, the nonlinear structural residual and
    tangent Jacobian are assembled and solved iteratively using an
    LM-regularized \textbf{least-squares procedure}, while aerodynamic loads are
    introduced through the work-consistent fluid--structure interface
    transfer. The computed geometry is sent to the fluid solver through the reverse transfer after time integration}
    \label{fig:solid_flowchart}
\end{figure}

\section{Coupling Methodology}
\label{sec:comp}

\texttt{VarFlExI} framework is formulated as a partitioned FSI problem in which the aerodynamic and structural solvers are advanced independently and exchange interface quantities at discrete coupling times. The following sections consist of four stages: specification of the fluid and structural boundary conditions in Section. (\ref{sec:boundary_conditions}), construction of the non-matching interface representations in Section. (\ref{sec:surface_discretization}), transfer of structural kinematics to the aerodynamic surface, and transfer of aerodynamic loads to the structural discretization in Section. (\ref{sec:interface_transfer}). The transfer is formulated to preserve the virtual work exchanged across the interface, as described in Section. (\ref{sec:work_transfer}).

\subsection{Boundary Conditions}
\label{sec:boundary_conditions}

The aerodynamic and structural subproblems are defined by their respective physical boundary conditions, together with the kinematic and dynamic conditions imposed at the fluid-structure interface. The aerodynamic boundary conditions determine the instantaneous flow solution on the deforming lifting surface, while the structural boundary conditions define the admissible deformation of the plate. The interface conditions are subsequently used to exchange the structural motion and aerodynamic loading between the two solvers.

\subsubsection{Aerodynamic Boundary Conditions}
\label{sec:aero_bcs}

The aerodynamic interface is treated as a deforming lifting surface, on which the normal
component of the relative flow velocity is constrained to vanish. At each aerodynamic control
point $\mathbf{x}_{\mathrm{cp},i}$, located at $\eta^c=3/4$, the kinematic boundary condition is
\begin{equation}
    \left[
    \mathbf{u}(\mathbf{x}_{\mathrm{cp},i},t)
    -
    \mathbf{u}_{\mathrm{kin},i}(t)
    \right]\cdot\mathbf{n}_i = 0,
    \label{eq:no_penetration}
\end{equation}
where $\mathbf{n}_i$ is the instantaneous unit normal and
$\mathbf{u}_{\mathrm{kin},i}$ is the local velocity of the deforming aerodynamic surface. Substitution of the velocity decomposition in Eq.~\eqref{eq:total_velocity} into Eq.~\eqref{eq:no_penetration} yields the algebraic system for the bound circulation distribution $\Gamma_i$.

The resulting unsteady bound circulation is coupled to wake shedding at the trailing edge.
Changes in the bound circulation are accompanied by the release of free vortex particles from
the trailing edge, $\eta^c=1$, thereby providing the discrete wake representation required by
the unsteady aerodynamic model.

\subsubsection{Structural Boundary Conditions}
\label{sec:struct_bcs}

The boundary of the structural plate domain $\partial \Omega_s$ is partitioned into a clamped root Dirichlet boundary $\Gamma_{s,\mathrm{root}} = \{(\eta^c, \eta^s) \mid \eta^s = 0\}$ at the wing-fuselage junction and a remaining free boundary $\Gamma_{s,\mathrm{free}} = \partial \Gamma_s \setminus \Gamma_{s,\mathrm{root}}$ along the leading edge, trailing edge, and wing tip. At the root ($\eta^s = 0$), rigid clamping is enforced by setting all five generalized kinematic degrees of freedom to zero: $
     u_0 = v_0 = w = \theta_x = \theta_y = 0 \quad \text{on } \Gamma_{s,\mathrm{root}}.
    \label{eq:clamped_bc}
$ Along the external edges ($\eta^c = 0$, $\eta^c = 1$, and $\eta^s = 1$), homogeneous Neumann conditions (traction-free) are applied for structural internal stresses. 

\subsection{Surface Discretization}
\label{sec:surface_discretization}

The aerodynamic and structural solvers employ independent, non-matching interface discretizations. The structural solver represents the deformable lifting surface using a finite-element discretization of the Reissner-Mindlin plate, whereas the aerodynamic solver represents the lifting surface using a two-dimensional Actuator Model.

The structural interface is defined on the plate mid-surface and discretized by the finite-element mesh described in Section. (\ref{sec:solid_model}). For the fluid-structure exchange, the structural displacement field is evaluated at $N_{\mathrm{comm}}$ spanwise communication stations parameterized by the normalized coordinates $(\eta^s,\eta^c)\in[0,1]^2$. At each spanwise communication station, the leading and trailing edges define the chordwise boundaries of the aerodynamic section, while the bound-vortex and control-point locations are defined at $\eta_{\mathrm{bv}}=1/4$ and $\eta_{\mathrm{cp}}=3/4$, respectively.
The leading and trailing-edge displacements provide the boundary data from which the intermediate aerodynamic locations are reconstructed. 

The present coupling framework supports both actuator-line and actuator-surface representations of the lifting surface. In the actuator-line formulation, the aerodynamic loading is evaluated at discrete actuator locations associated with the lifting surface. The actuator-surface formulation provides an alternative within the same framework when a distributed representation of bound vorticity is desirable as discussed in Section. (\ref{sec:rVPM_coupling}) \citep{alvarez2022flowunsteady,ASM}.

\subsection{Common Interface Parametrization}
\label{sec:interface_transfer}

\begin{algorithm}[t]
\caption{Parametric interpolation between non-matching interface discretizations}
\label{alg:parametric_interpolation}
\begin{algorithmic}[1]
\State \textbf{Input:} Interface quantity $\mathbf r(\eta_s^{\mathrm{in}},\eta_c^{\mathrm{in}})$,
$\eta_s^{\mathrm{in}},\eta_c^{\mathrm{in}},
\eta_s^{\mathrm{out}},\eta_c^{\mathrm{out}}$
\State \textbf{Output:} Interpolated quantity $\mathbf r_c(\eta_s^{\mathrm{out}},\eta_c^{\mathrm{out}})$

\State $\eta_s^{\mathrm{in}},\eta_c^{\mathrm{in}},
\eta_s^{\mathrm{out}},\eta_c^{\mathrm{out}}
\gets \texttt{as\_eta\_array}(\cdot)$

\For{$j=1,\ldots,N_c^{\mathrm{in}}$}
    \State Do chord-wise interpolation, 
    $\widetilde{\mathbf r}_{:,j}
    \gets
    \texttt{interp\_profile}
    \left(
    \eta_s^{\mathrm{in}},
    \mathbf r_{:,j},
    \eta_s^{\mathrm{out}}
    \right)$ 
\EndFor

\For{$i=1,\ldots,N_s^{\mathrm{out}}$}
    \State Do span-wise interpolation
    $\mathbf r_{c,i,:}
    \gets
    \texttt{interp\_profile}
    \left(
    \eta_c^{\mathrm{in}},
    \widetilde{\mathbf r}_{i,:},
    \eta_c^{\mathrm{out}}
    \right)$
\EndFor

\State \textbf{where} $\texttt{interp\_profile}(x_{\mathrm{in}},
\mathbf r_{\mathrm{in}},x_{\mathrm{out}})$ is given by
\[
\mathbf r(x)
=
(1-\lambda)\mathbf r^{(k)}
+
\lambda\mathbf r^{(k+1)},
\lambda=
\frac{x-x^{(k)}}
{x^{(k+1)}-x^{(k)}} \qquad
 \text{for } 
x^{(k)}\leq x\leq x^{(k+1)}\]

\State \Return $\mathbf r_c(\eta_s^{\mathrm{out}},\eta_c^{\mathrm{out}})$
\end{algorithmic}
\end{algorithm}

The fluid and structural solvers employ independent, non-conforming
discretizations and represent the exchanged quantities in different solver-specific spaces. A common normalized parameterization is therefore introduced as an intermediate interface representation. Let $a\in\{f,s\}$ denote the aerodynamic and structural solvers, respectively,
and let $\mathbf{r}_a$ denote a quantity defined on the corresponding
solver-specific interface. The interface points are parameterized by the
normalized spanwise coordinate $\eta_s\in[0,1]$ and normalized chordwise
coordinate $\eta_c\in[0,1]$, where $\eta_s=0$ and $\eta_s=1$ denote the root
and tip, respectively, and $\eta_c=0$ and $\eta_c=1$ denote the leading and
trailing edges. The solver-specific discrete field is written as
\begin{equation}
    \mathbf{r}_a
    =
    \left\{
    \mathbf{r}_{a,ij}
    \right\},
    \qquad
    \mathbf{r}_{a,ij}
    =
    \mathbf{r}_a
    \left(
    \eta_{s,i}^{a},
    \eta_{c,j}^{a}
    \right),
    \qquad a\in\{f,s\},
\end{equation}
defined on the solver-specific interface grid with $N_s^a$ spanwise and
$N_c^a$ chordwise locations. The common coupling interface is parameterized
by $\eta_s\in \left\{ \eta_{s,i}^{c} \right\}_{i=1}^{N_s^c}, \eta_c\in \left\{\eta_{c,j}^{c} \right\}_{j=1}^{N_c^c},$ where the superscript $c$ denotes the common interface. Since the solver-specific discretizations generally differ from the common coupling grid, $\mathbf{r}_a$ is first resampled in the normalized parameter space $(\eta_s,\eta_c)$ before the subsequent solver-to-solver transfer operation.

For a quantity defined at
$\left(\eta_{s,i}^{a},\eta_{c,j}^{a}\right)$, the spanwise interpolation is
first performed according to \eqref{eq:interpolation}, followed by interpolation in the chordwise direction,

\begin{equation}
    \widetilde{\mathbf{r}}_{a,ij}
    =
    \mathcal{I}_{\eta_s}
    \left[
    \mathbf{r}_{a}
    \left(
    \eta_s,\eta_{c,j}^{a}
    \right)
    \right]_{\eta_s=\eta_{s,i}^{c}},
    \qquad
    \mathbf{r}_{c,ij}
    =
    \mathcal{I}_{\eta_c}
    \left[
    \widetilde{\mathbf{r}}_{a}
    \left(
    \eta_{s,i}^{c},\eta_c
    \right)
    \right]_{\eta_c=\eta_{c,j}^{c}}.
    \label{eq:interpolation}
\end{equation}

Accordingly, the solver-to-common resampling operation is written as
\begin{equation}
    \mathbf{r}_c
    =
    \mathcal{I}_{\eta_c}
    \circ
    \mathcal{I}_{\eta_s}
    (\mathbf{r}_a),
    \qquad a\in\{f,s\}.
    \label{eq:common_interface_resampling}
\end{equation}

For the fluid-to-structure load transfer, the aerodynamic solver provides
the discrete force field $\mathbf{F}_f$, which is first resampled onto the
common interface according to Eq.~\eqref{eq:common_interface_resampling}.
The resulting common-interface force representation is then transferred
to the structural finite-element space through the force transfer operator $\mathcal{T}^{F}_{c\rightarrow s}$,
\begin{equation}
    \mathbf{F}_s
    =
    \mathcal{T}^{F}_{c\rightarrow s}
    \left[
    \mathcal{I}_{f\rightarrow c}
    \left(
    \mathbf{F}_f
    \right)
    \right].
    \label{eq:forward_force_transfer}
\end{equation}
Thus, the parametric resampling establishes the geometric correspondence between the two discretizations, while
$\mathcal{T}^{F}_{c\rightarrow s}$ performs the subsequent work-conservative
load transfer.

The reverse transfer of structural kinematics follows the same
common-interface framework, but acts on the geometric representation of the
aerodynamic surface. The structural displacement field $\mathbf{u}_s$ is first interpolated onto the common interface, and then mapped to the aerodynamic geometric locations using a geometry-transfer operator.

\begin{equation}
    \mathbf{u}^{g}_f
    =
    \mathcal{T}^{G}_{c\rightarrow f}
    \left[
    \mathcal{I}_{s\rightarrow c}
    \left(
    \mathbf{u}_s
    \right)
    \right]
    \label{eq:solid_to_common_geometry}
\end{equation}

The geometry representation of the fluid solver,
$\mathbf{u}^{g}_f =
\begin{bmatrix}
\mathbf{u}_{LE} & \mathbf{u}_{TE}
\end{bmatrix}^{T}$,
is subsequently used to reconstruct the aerodynamic locations required by the flow solver. We denote this reconstruction by the linear interpolation
operator $\mathcal{R}$,
\begin{equation}
    \mathbf{u}_{b}
    =
    \mathcal{R}_{b}
    \left(
    \mathbf{u}^{g}_f
    \right),
    \qquad
    b\in\{CP,BV\},
    \qquad
    \eta_{CP}=\frac{3}{4},
    \quad
    \eta_{BV}=\frac{1}{4},
    \label{eq:geometry_reconstruction}
\end{equation}
where $\mathcal{R}_{b}$ denotes linear interpolation between the leading and trailing-edge displacements at the corresponding normalized chordwise location $\eta_b$. The following section addresses the enforcement of work conservation during the interface exchanges.

\begin{figure}[pos=h!]
    \centering
    \includegraphics[width=0.9\linewidth]{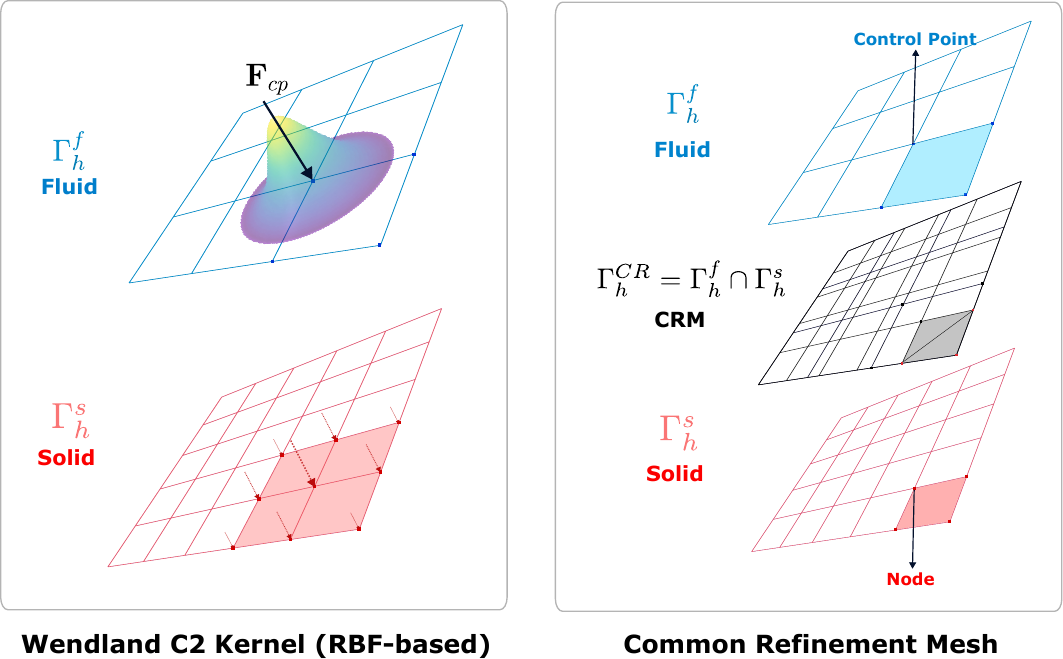}
\caption{Conservative interface work-transfer schemes for
non-matching fluid and structural discretizations. The aerodynamic
load field is transferred to the structural representation using
(left) a \textbf{Wendland-$C^2$ RBF}-based interpolation and (right) a
\textbf{Common-Refinement mesh} (CRM) construction, with the corresponding
fluid control points and structural nodes illustrated.}
    \label{fig:scoupling_flowchart}
\end{figure}

\subsection{Work Conservative Interface Transfer}
\label{sec:work_transfer}

In partitioned FSI problems, the fluid and structural solvers exchange interface quantities through independent numerical discretizations. A physically consistent coupling requires that the transfer of interface loads does not artificially create or dissipate energy. This is commonly enforced through the principle of work conservation \citep{farhat_workcons}, whereby the virtual work performed by the aerodynamic loads is preserved during the transfer from the meshless aerodynamic solver to the structural discretization. Mathematically, this condition is given by:

\begin{equation}
\delta W_f=\delta W_s,
\end{equation}

where $\delta W_f$ and $\delta W_s$ denote the virtual work evaluated on the fluid and structural interfaces, respectively. The aerodynamic and structural solvers employed in the present framework utilize fundamentally different spatial discretizations. Consequently, the aerodynamic force space and the structural displacement space are non-matching, requiring a consistent transfer operator for the exchange of interface quantities \citep{SLATTERY2016164}.

The following development first establishes the work-conservation requirement in a generalized setting. Subsequently, we consider two interpolation strategies for constructing the interface operator, namely a compact supported Radial Basis Function \citep{rbf_workcons, aeroelasticmdo}, and a common refinement operator \citep{crm_initial,crm}. Without loss of generality, let $\Gamma_f$ and $\Gamma_s$ denote the fluid and structural interface meshes, respectively. The continuous traction fields defined on the two interfaces are approximated independently using their respective interpolation functions as,

\begin{equation}
\mathbf{t}^{\,f}(\mathbf{x})
\approx
\sum_{k=1}^{m_f}
N_i^{\,f}(\mathbf{x})
\,
\tilde{\mathbf{t}}_i^{\,f}, \quad 
\mathbf{t}^{\,s}(\mathbf{x})
\approx
\sum_{k=1}^{m_s}
N_j^{\,s}(\mathbf{x})
\,
\tilde{\mathbf{t}}_j^{\,s},
\label{eq:traction_interp}
\end{equation}

where $m_f$ and $m_s$ denote the number of fluid and structural interface nodes (or coupling points), respectively, $N_i^{\,f}$ and $N_j^{\,s}$ represent the corresponding interpolation functions, and $\tilde{\mathbf{t}}_i^{\,f}$ and $\tilde{\mathbf{t}}_j^{\,s}$ are the unknown nodal traction vectors. Since the two interface discretizations are generally non-coincident, the transferred structural traction field cannot satisfy the traction equilibrium condition pointwise. Instead, the residual $
\mathbf{r}
=
\mathbf{t}^{\,s}
-
\mathbf{t}^{\,f},
$ is required to be orthogonal to the structural approximation space in the Galerkin sense, given by $\int_{\Gamma}
N_i^{\,s}
\mathbf r
\,d\Gamma
=
0,
i=1,\ldots,m_s$. Hence, this yields

\begin{equation}
\int_{\Gamma_s}
N_i^{\,s}
\mathbf{t}^{\,s}
\,d\Gamma_s
=
\int_{\Gamma_s}
N_i^{\,s}
\mathbf{t}^{\,f}
\,d\Gamma_s,
\qquad
i=1,\ldots,m_s.
\label{eq:galerkin_proj}
\end{equation}

On substituting Eq. (\ref{eq:traction_interp}) into Eq. (\ref{eq:galerkin_proj}), we obtain

\begin{equation}
\sum_{k=1}^{m_s}
\left(
\int_{\Gamma}
N_i^{\,s}
N_k^{\,s}
\,d\Gamma
\right)
\tilde{\mathbf{t}}_j^{\,s}
=
\sum_{k=1}^{m_f}
\left(
\int_{\Gamma}
N_i^{\,s}
N_k^{\,f}
\,d\Gamma
\right)
\tilde{\mathbf{t}}_k^{\,f}.
\label{eq:general_projection}
\end{equation}

Equation (\ref{eq:general_projection}) constitutes the general Galerkin projection governing the conservative transfer of interface traction between non-matching discretizations. Note that, in the following sections, the aerodynamic force $\mathbf{F}_f$ and the structural geometry $\mathbf{u}_s$ are the quantities represented in the common interface after interpolation. For convenience, we denote them by $\mathbf{F}_f$ and $\mathbf{u}_s$, rather than the notation used in Section. (\ref{sec:interface_transfer})

\subsubsection{Dual-Operator Requirement for Multi-representative discretization}
\label{sec:bidirectional_in_rvpm}

The coupling between the meshless \textit{r-VPM} aerodynamic solver and the mesh-based structural solver involves distinct aerodynamic representations
for surface geometry and aerodynamic quantities. The structural interface is defined by the finite-element nodes, whereas the aerodynamic surface is
defined by its surface vertices, from which the bound surface elements are constructed, making it a 'multi'-- 
 representative discretization. The no-penetration condition is enforced at control points, and
the local velocity used in the aerodynamic force evaluation is likewise obtained at these points, with the resulting load associated with the corresponding bound surface elements. Thus, the aerodynamic surface geometry and loading enter the coupling through different discrete representations.

A direct transpose of the structural-to-aerodynamic displacement operator as in \citep{aeroelasticmdo,PARENTEAU2020103146} is therefore not sufficient to reconstruct the deformed aerodynamic surface. The displacement transfer is instead formulated through the common interface, using an RBF-based geometry-transfer operator followed by reconstruction of the required aerodynamic locations. This distinct geometric representation does not preclude work conservation, since conservation is imposed on the complete force and displacement mappings rather than by requiring the
individual transfer operators to be transpose pairs.

Let $\mathbf{F}_c$ denote the aerodynamic force after parametric resampling to
the common interface and $\mathbf{u}_c$ the corresponding structural
displacement. The aerodynamic displacement at the force-evaluation locations
is obtained as

\begin{equation}
    \mathbf{u}_f
    =
    \mathcal{R}
    \mathcal{T}^{G}_{c\rightarrow f}
    \mathcal{I}_{s\rightarrow c}
    \mathbf{u}_s ,
    \label{eq:aero_displacement_map}
\end{equation}

where $\mathcal{R}$ reconstructs the required aerodynamic displacement
locations from the transferred surface geometry. The structural force is
obtained from

\begin{equation}
    \mathbf{F}_s
    =
    \mathcal{T}^{F}_{c\rightarrow s}
    \mathcal{I}_{f\rightarrow c}
    \mathbf{F}_f ,
    \label{eq:solid_force_map}
\end{equation}

The corresponding virtual works are
\begin{equation}
    \delta W_f
    =
    \left[
    \mathcal{R}
    \mathcal{T}^{G}_{c\rightarrow f}
    \mathcal{I}_{s\rightarrow c}
    \delta\mathbf{u}_s
    \right]^T
    \mathbf{F}_f ,
    \qquad
    \delta W_s
    =
    \delta\mathbf{u}_s^T
    \mathcal{T}^{F}_{c\rightarrow s}
    \mathcal{I}_{f\rightarrow c}
    \mathbf{F}_f ,
    \qquad
    \delta W_f=\delta W_s .
    \label{eq:virtual_work}
\end{equation}

Defining the complete force and displacement mappings as
\begin{equation}
    \mathcal{H}_F
    =
    \mathcal{T}^{F}_{c\rightarrow s}
    \mathcal{I}_{f\rightarrow c},
    \qquad
    \mathcal{H}_G
    =
    \mathcal{R}
    \mathcal{T}^{G}_{c\rightarrow f}
    \mathcal{I}_{s\rightarrow c},
    \qquad
    \mathcal{H}_F=\mathcal{H}_G^T ,
    \label{eq:composite_maps}
\end{equation}

This condition ensures that the forward force and reverse displacement mappings are work-conjugate, thereby yielding identical virtual work on the
aerodynamic and structural sides for arbitrary compatible displacements. Thus, the transpose relation applies to the complete work-conjugate mappings,
rather than directly between $\mathcal{T}^{F}_{c\rightarrow s}$ and
$\mathcal{T}^{G}_{c\rightarrow f}$.

\subsubsection{Compactly Supported Wendland Kernel Operator} 

Unlike projection methods that require explicit evaluation of coupling integrals over the interface, a mesh-free \textit{Radial Basis Function} (RBF) operator using compactly supported kernels \citep{SLATTERY2016164} can be employed to construct a continuous interpolation operator directly from geometric locations of the interface nodes, making the transfer procedure independent of the underlying mesh connectivity \citep{rbf_workcons, aeroelasticmdo}. Given the force field parametrization, Section. (\ref{sec:interface_transfer}), the force grid is defined as $(x_i^f,F_i^f)$ and $(x_j^s,F_j^s)$, the interpolation weights are then constructed according to the spatial separation between the communication point and the neighboring structural nodes. The transfer uses a compact-support Wendland $C^2$ Radial Basis Kernel, defined as

\begin{equation}
\phi(r)
=
\begin{cases}
(1-q)^4(4q+1), & q<1,\\
0, & q\ge1,
\end{cases}
\end{equation}

where
$
q=\frac{r}{R}.
$ Here $r$ denotes the Euclidean distance between the interface points, and $R$ is the compact support radius controlling the extent of the interpolation. 
For a communication point $\mathbf{x}_i^f$, the raw interpolation weights are computed as
$w_{ij}^{'}
=
\phi\left(
\|\mathbf{x}_i^f-\mathbf{x}_j^s\|_2
\right)
$. To ensure local conservation of force magnitude, the weights are normalized according to
\begin{equation}
w_{ij}
=
\frac{
w_{ij}^{'}
}{
\sum_{k \in \mathcal{N}_i}
w_{ik}^{'}
},
\end{equation}
where $\mathcal{N}_i$ denotes the local neighborhood associated with communication point $i$, defined as $\mathcal{N}_i=\{x_j^s:||x_i^f-x_j^s||<R\}$. The structural nodal forces and therefore $\mathbf{F}_{\mathrm{eff}}$ are subsequently obtained by weighted accumulation
\begin{equation}
\mathbf{F}_j^s
=
\sum_{i=1}^{N_f}
w_{ij}\mathbf{F}_i^f.
\label{eq:rbf_transfer}
\end{equation}

Equation~(\ref{eq:rbf_transfer}) may be written compactly as

\begin{equation}
\mathbf F_s
=
H_f^T
\mathbf F_f,
\end{equation}

where the transfer matrix $H_f$ is assembled from the normalized Wendland kernel weights. The displacement transfer, implemented separately as another Wendland Kernel operator, $
\mathbf u_f
=
H_s^T
\mathbf u_s,$ preserves the virtual work exchanged across the fluid-structure interface.

\subsubsection{Common Refinement Mesh-based Interface Transfer}
\label{sec:crm}

 RBF-based work transfer operators are able to provide an efficient and mesh agnostic transfer strategy, their conservative formulation requires a polynomial augmentation to exactly reproduce rigid-body motions \citep{flyer_rbf}. Furthermore in our implementation of the Wendland $C^2$ kernel, the quality of transfer depends on the choice of the support radius, and the distribution of interface nodes. While compact support significantly reduces computational costs by constructing sparse transfer matrices, they do not account for the overlap between the fluid and solid structural interface, resulting in reduced accuracy of transferred quantities for highly non-uniform meshes, large mesh-ratio differences, and complex interfaces \citep{SLATTERY2016164}.

Eq.~\eqref{eq:general_projection} establishes the conservative Galerkin projection between the aerodynamic and structural interface. Although it provides a continuous weak form of the transfer problem, its numerical evaluation becomes non-trivial for non-matching interface meshes, especially for the term $N_i^{\,s},
N_k^{\,f}$ in the integral \eqref{eq:general_projection}. To overcome this, we extend the idea of a \textit{Common Refinement Operator} \citep{crm_initial,crm,SLATTERY2016164} to a discretized fluid domain-meshed solid domain panel case. We construct a geometric auxiliary overlay by intersecting the aerodynamic panels associated with their control points with the finite elements of the structural discretization. The resulting overlap regions provide a common integration domain on which aerodynamic panel forces are conservatively projected onto the structural finite element basis while preserving the Galerkin weak form. It partitions the interface into a set of overlap regions formed by the intersection of aerodynamic panels with structural finite elements, 

\begin{equation}
\Gamma
=
\bigcup_{m=1}^{N_{cr}}
\Omega_m^{cr},
\label{eq:crm_partition}
\end{equation}

where $\Omega_m^{cr}$ denotes the $m^{\mathrm{th}}$ overlap polygon and
$N_{cr}$ is the total number of overlap regions. Hence the Galerkin projection (Eq. (\ref{eq:galerkin_proj})) is therefore evaluated over the common refinement surface as $
\mathbf{f}^{\,s}
=
\sum_{m=1}^{N_{cr}}
\int_{\Omega_m^{cr}}
\mathbf N^{\,s}
\mathbf t^{\,f}
\,d\Gamma.
\label{eq:crm_integral}
$ Since \texttt{FLOWUnsteady} associates a single resultant force $\mathbf F_p^{\,f}$ with each panel, we assume that the traction $t^f$ is constant for each of the panel 

\begin{equation}
\mathbf t_p^{\,f}
=
\frac{\mathbf F_p^{\,f}}
{A_p},
\label{eq:panel_traction}
\end{equation}

where $A_p$ denotes the panel area. Now the overlap integrals can be assembled into a global transfer operator $
\mathbf C
=
\left[
C_{ip}
\right],$ whose entries are given by

\begin{equation}
C_{ip}
=
\sum_{\Omega_m^{cr}\subset\Gamma_p}
\int_{\Omega_m^{cr}}
N_i^{\,s}
\,d\Gamma,
\label{eq:Cmatrix}
\end{equation}

where $\Gamma_p$ denotes the $p^{\mathrm{th}}$ aerodynamic panel. The equivalent structural nodal force vector is therefore obtained as

\begin{equation}
\mathbf F_{s}
=
\mathbf C
\mathbf t^{\,f}
\label{eq:crm_final}
\end{equation}

\begin{algorithm}
\caption{Common-Refinement Mesh-based Conservative Interface Transfer}
\label{alg:crm_transfer}
\begin{algorithmic}[1]
\State \textbf{Input:} Structural FE mesh $\mathcal{M}_s$, fluid coupling stations $\{\eta_i\}$,
fluid panel geometry $\{P_i\}$, structural displacement field $\mathbf{u}_s$,
fluid panel forces $\mathbf{F}_f$
\State \textbf{Output:} Structural nodal forces $\mathbf{F}_s$,
fluid-interface displacements $\mathbf{u}_f$, (optional)  $\mathbf{C} \in \mathbb{R}^{N_{{dof},s} \times N_f}$

\State Construct panel boundaries, $(y_i,y_{i+1}) \gets \texttt{panel\_boundary}(\eta_i)$
\State Construct fluid panels, $P_i \gets \texttt{panel\_polygon\_for\_span\_interval}(y_i,y_{i+1})$
\State Pre-compute panel areas, $A_i \gets 
\texttt{polygon\_area}(P_i)$
\State \texttt{Initialize} $\mathbf{C} \gets
\mathbf{0}$
\State \texttt{Initialize} $\mathbf{A} \gets \texttt{diag}(A_i)\succeq 0$
\For{each structural element $T_j$}
    \For{each fluid panel $P_i$}
\If{$\texttt{bbox}(T_j)\cap\texttt{bbox}(P_i)=\emptyset$}
    \State \textbf{continue}
\EndIf
        \State Compute geometric overlap as $\Omega_{ij} \gets T_j \cap P_i$
        \If{$\operatorname{\texttt{area}}(\Omega_{ij}) = 0$}
            \State \textbf{continue}
        \EndIf
        \State \texttt{triangulate}($\Omega_{ij}$)
        \State Integrate structural basis over $\Omega_{ij}$
        \State Update $\mathbf{C}_{ji} \gets \mathbf{C}_{ji}+\displaystyle\int_{\Omega_{ij}} N_j\,\mathrm{d}A$
    \EndFor
\EndFor

\State $\mathbf{f}_{f,i} \gets \mathbf{F}_{f,i}/A_i$
\State $\mathbf{F}_s \gets \mathbf{C}\mathbf{f}_f$
\State $\mathbf{u}_f \gets \mathbf{A}^{-1}\mathbf{C}^{T}\mathbf{u}_s$
\State \Return $\mathbf{F}_s,\mathbf{u}_f$
\end{algorithmic}
\end{algorithm}

The resulting partitioned framework is next evaluated through a sequence of numerical experiments and validation studies designed to quantify coupling accuracy, energy consistency, and spatiotemporal convergence over a range of flow and structural conditions.

\section{Experiments}
\label{sec:expt}

This section assesses the accuracy, energy consistency, and spatiotemporal convergence of the partitioned \texttt{VarFlExI} framework. The baseline test case considers an untapered, flexible cantilever plate wing impulsively immersed in an incompressible freestream. Unless otherwise specified in parametric sweep studies, the physical, structural, aerodynamic, and coupling parameters are held fixed at their nominal baseline values listed in Table~\ref{tab:simulation_parameters}.

\begin{table*}[t]
\centering
\caption{Physical, structural, and nominal numerical parameters used in the sensitivity studies.}
\label{tab:simulation_parameters}

\begin{tabularx}{\textwidth}{l X l X}
\toprule
\textbf{Notation} & \textbf{Nominal Value} &
\textbf{Notation} & \textbf{Nominal Value} \\
\midrule

$U_\infty$ & $8.0\text{ m/s}$
& $b$ & $0.80\text{ m}$
\\

$\rho_f$ & $1.0\text{ kg/m}^3$
& $c_{\mathrm{root}},\,c_{\mathrm{tip}}$
& $0.12\text{ m}$ ($AR=6.67$)
\\

$\nu_f$ & $1.0\times10^{-6}\text{ m}^2/\text{s}$
& $h/c$ & $0.12$ ($h=0.0144\text{ m}$)
\\

$\alpha$ & $8.0^\circ$
& & 
\\
\midrule

$E$ & $3.0\times10^{10}\text{ Pa}$ ($30\text{ GPa}$)
& $\eta_m$ & $0.8\text{ s}^{-1}$
\\

$\nu_s$ & $0.35$
& $\eta_k$ & $1.0\times10^{-4}\text{ s}$
\\

$\rho_s$ & $1600\text{ kg/m}^3$
& $\kappa_{\mathrm{shear}}$ & $0.833$ ($5/6$)
\\

$\alpha_m,\,\alpha_f$ & $0.1,\,0.2$
& & 
\\
\midrule

$N_x\times N_y$ & $50\times150$ elements
& $\Delta t$ & $0.001\text{ s}$
\\

$N_{\mathrm{span}}\times N_{\mathrm{chord}}$
& $80\times1$ panels
& $t_{\mathrm{tot}}$ & $5.0\text{ s}$
\\

$N_p^{(k)}-N_p^{(k-1)}$
& $1\text{particle/step}$
& \text{Transfer Method} & \text{Common Refinement Mesh}
\\

\bottomrule
\end{tabularx}
\end{table*}

To systematically evaluate solver independence, parametric sweeps are performed across the temporal, spatial, and wake discretization resolutions as summarized in Table~\ref{tab:sweep}.


\subsection{Work Conservation Metric}
\label{sec:work_conservation_comparison}

To evaluate the virtual-work consistency of the interface load and displacement mapping, the performance of the Common Refinement Mesh operator is benchmarked
against a Compactly Supported Radial Basis Function (RBF) transfer scheme utilizing a Wendland $C^2$ kernel, as described in Section~\ref{sec:work_transfer}. The test case corresponds to the impulsive flow start of the flexible cantilever plate wing, with parameters discussed in the following Table \ref{tab:simulation_parameters}. The energy consistency of the interface transfer is quantified by the instantaneous relative virtual work conservation error, given by

\begin{equation}
  \epsilon_W(t) = \frac{\left| \mathbf{F}_s(t)^T \delta \mathbf{u}_s(t) - \mathbf{F}_f(t)^T \delta \mathbf{u}_f(t) \right|}{\left| \mathbf{F}_f(t)^T \delta \mathbf{u}_f(t) \right| + \epsilon_{\mathrm{reg}}},
  \label{eq:work_error_metric}
\end{equation}

where $\mathbf F_f$ and $\delta\mathbf u_f$ denote the aerodynamic force vector and corresponding virtual displacement vector in the work-conjugate aerodynamic representation, while $\mathbf F_s$ and $\delta\mathbf u_s$ denote the corresponding structural quantities, and $\epsilon_{\mathrm{reg}} = 10^{-16}$ is a regularization constant to prevent undefined cases.


\begin{figure}[pos=h!]
    \centering
    \includegraphics[width=0.75\linewidth]{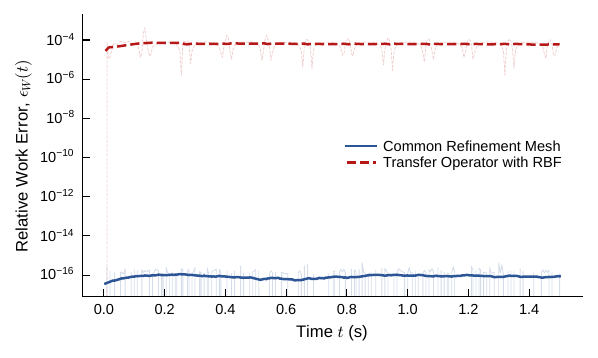}
    \caption{Comparison of the Common Refinement Mesh (CRM) and Compactly Supported Wendland $C^2$ RBF transfer operators for conservative interface coupling. Simulation parameters: $N_x \times N_y = 50 \times 150$, $N_{\mathrm{span}} \times N_{\mathrm{chord}} = 80 \times 1$, $\Delta t = 0.001\text{ s}$, RBF support radius $r_0 = 0.08$, $24$ nearest neighbors, and maximum absolute force threshold $5.0\times 10^3\text{ N}$.}
    \label{fig:work_cons_comparison}
\end{figure}

Figure \ref{fig:work_cons_comparison} presents the time evolution of the relative work conservation error $\epsilon_W(t)$ over the complete simulation duration. Because the CRM construction provides the work-equivalent force and displacement mappings across the non-matching interfaces, the resulting composite operators satisfy the discrete adjoint relation required for virtual work conservation. Consequently,
$\mathbf{F}_s^T\delta\mathbf{u}_s = \mathbf{F}_f^T\delta\mathbf{u}_f$ up to numerical roundoff. As shown by the solid blue curve, the relative work error remains bounded between $10^{-16}$ and $10^{-15}$ across all time steps, preserving interface virtual work to strict machine precision. During the initial impulsive startup, $t \le 0.5\text{ s}$, large spatial force gradients cause the RBF work error to peak at $\epsilon_W \approx 10^{-2}$. As the flow and structural transients settle ($t > 1.5\text{ s}$), the RBF error stabilizes to an asymptotic plateau of $\epsilon_W \approx 6.0 \times 10^{-5}$. The CRM-based formulation thus achieves a virtual work preservation accuracy that is approximately $10$ to $11$ orders of magnitude superior to the standard RBF operator, eliminating artificial energy generation or dissipation at the non-matching discrete interface.





\subsection{Heaving Experiments}
\label{sec:validation}


To validate the two-way aeroelastic coupling and the associated kinematic interface transfer, the present framework is benchmarked against the water-tunnel measurements of Heathcote et al. \citep{validation}. The experiments considered a rectangular flexible wing undergoing
pure harmonic heaving in a free-surface closed-loop water tunnel. The wing was mounted vertically with one end attached to a horizontal mechanical shaker, which prescribed the root motion, while the remaining span was free to deform under the hydrodynamic loading. The experiments investigated the effect of spanwise flexibility on the resulting thrust, lift, and propulsive efficiency over a range of Reynolds numbers and oscillation frequencies. A two-component force balance was used to measure the streamwise and transverse forces on the wing, while Particle Image Velocimetry (PIV) was employed to characterize the resulting vortical flow field \citep{validation}. The experiments used a constant root-heave amplitude $a_{\mathrm{ROOT}}/c=0.175$, with the unsteadiness characterized by the Reynolds number
$Re=\rho_f U_\infty c/\mu$ and Garrick frequency
$k_G=\pi f c/U_\infty$ \citep{validation}.

In the present numerical reproduction, the experimental root actuation is imposed as a prescribed transverse displacement of the clamped structural root, while the in-plane translations and rotations at the root remain constrained. To avoid introducing an artificial impulsive loading at the beginning of the simulation, the harmonic displacement is smoothly ramped over the first three oscillation periods before reaching its prescribed steady amplitude. The structural response away from the driven root is therefore obtained from the coupled \textit{FSI} rather than prescribed kinematically, allowing the spanwise deformation and its feedback on the aerodynamic loading to emerge from the coupled solution.

For the heaving validation, the clamped-root boundary condition is retained for the in-plane displacements and rotations, while the transverse displacement at the root is prescribed as a harmonic heave, $
w_{\mathrm{root}}(t)
=
A\cos(\omega t),
$ where $A=0.175c$. To avoid an impulsive start, the prescribed motion is smoothly ramped over three oscillation periods,
\begin{equation}
w_{\mathrm{root}}(t)
=
\begin{cases}
A\sin^2\left(\dfrac{\pi t}{2T_{\mathrm{ramp}}}\right)
\cos(\omega t),
& t<T_{\mathrm{ramp}},\\[6pt]
A\cos(\omega t),
& t\geq T_{\mathrm{ramp}},
\end{cases}
\qquad
T_{\mathrm{ramp}}=\dfrac{3}{f}.
\end{equation}
The remaining root degrees of freedom are constrained according to $
u_x=u_y=0,
\theta_x=\theta_y=0
\text{ on } \Gamma_{\mathrm{root}}.
$ The kinematic regime is characterized by the reduced frequency, $
k_G = \frac{\pi f c}{U_\infty},
\label{eq:reduced_freq}
$ where $f$ is the prescribed heaving frequency, $c$ is the chord length, and $U_\infty$ is the freestream velocity. The corresponding period of one complete heaving cycle is
\begin{equation}
T = \frac{1}{f}
  = \frac{\pi c}{k_G U_\infty}.
\end{equation}
To assess the coupled solver across different structural flexibilities, two configurations considered in \citep{validation} are simulated: a flexible plate with Young's modulus $E=3.4\times10^{9}\,\mathrm{Pa}$ and a stiffer plate with $E=2.6\times10^{10}\,\mathrm{Pa}$. For both configurations, the root is prescribed to undergo harmonic heaving, with the motion smoothly ramped over the initial three oscillation periods to minimize startup transients. A total of 80 heaving cycles are simulated using a coupling time step of $\Delta t=5.0\times10^{-3}\,\mathrm{s}$. The initial transient response associated with wake development is discarded, and the aerodynamic and structural responses are evaluated over the subsequent statistically periodic cycles.

\begin{table}[htbp]
\centering
\caption{Parameters for validation against Heathcote et al. \citep{validation}.}
\label{tab:val_table}
\renewcommand{\arraystretch}{1.25}
\begin{tabular}{|c|l|c|l|}
\hline
\textbf{Symbol} & \textbf{Description} & \textbf{Symbol} & \textbf{Description} \\
\hline
$Re$          & $30\,000$                 & $\nu$           & $0.35$                 \\
$E$           & $3.4 \times 10^9$ / $2.6 \times 10^{10}\text{ Pa}$  & $\text{span}$   & $0.30\text{ m}$               \\
$\rho_f$        & $998\text{ kg/m}^3$          & $\text{chord}$  & $0.10\text{ m}$               \\
$U_{\infty}$  & $0.30\text{ m/s}$              & $\text{cycles}$   & $80$                   \\
$k_G$         & $1.82$ ($ \in 
(0.3 - 1.9
)$)                  & $\Delta t$            & $5.0 \times 10^{-3}\text{ s}$ \\
\hline
\end{tabular}
\end{table}

\begin{figure}[pos=h!]
    \centering
    \begin{subfigure}[b]{0.48\textwidth}
        \centering
        \includegraphics[width=\textwidth, page=1]{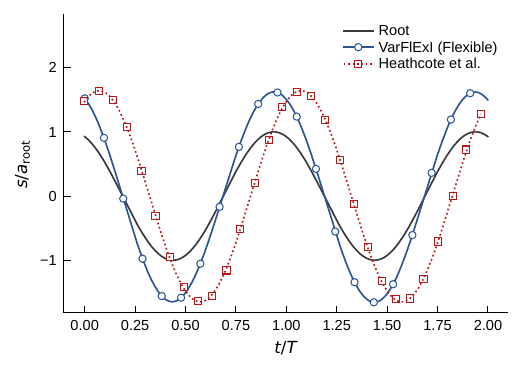}
        \caption{Flexible plate ($E = 3.4 \times 10^9\text{ Pa}$).}
        \label{fig:fig4_flex}
    \end{subfigure}
    \hfill
    \begin{subfigure}[b]{0.48\textwidth}
        \centering
        \includegraphics[width=\textwidth, page=1]{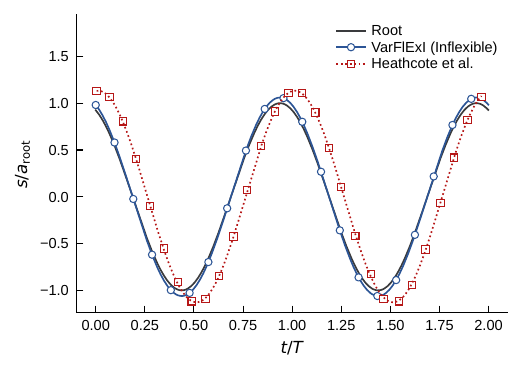}
        \caption{Less flexible plate ($E = 2.6 \times 10^{10}\text{ Pa}$).}
        \label{fig:fig4_inflex}
    \end{subfigure}
    \caption{Time history of the normalized tip displacement $s/a_{\mathrm{root}}$ over normalized time $t/T$ for two full heaving cycles at $k_G = 1.82$, validated against Heathcote et al. \citep{validation}. The blue curve represents the \texttt{VarFlExI} solution, the red one indicates the experimental result from \citep{validation}, and the black curve is the root heaving.}
    \label{fig:fig4}
\end{figure}

\begin{figure}[pos=h!]
    \centering
    \begin{subfigure}[b]{0.48\textwidth}
        \centering
        \includegraphics[width=\textwidth, page=1]{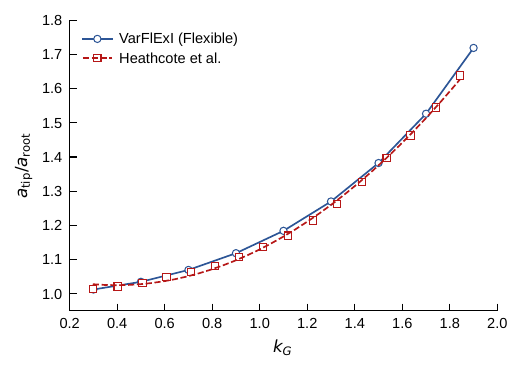}
        \caption{Flexible plate ($E = 3.4 \times 10^9\text{ Pa}$).}
        \label{fig:fig5a_flex}
    \end{subfigure}
    \hfill
    \begin{subfigure}[b]{0.48\textwidth}
        \centering
        \includegraphics[width=\textwidth, page=1]{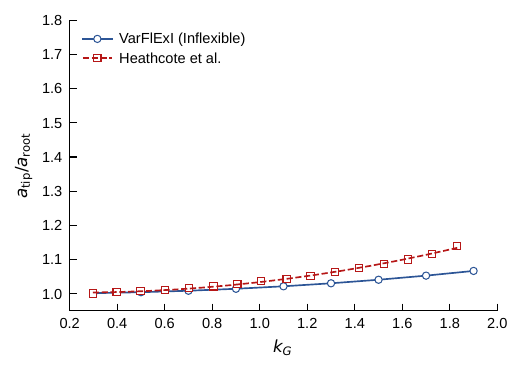}
        \caption{Less flexible plate ($E = 2.6 \times 10^{10}\text{ Pa}$).}
        \label{fig:fig5a_inflex}
    \end{subfigure}
    \caption{Normalized peak tip displacement amplitude $a_{\mathrm{tip}}/a_{\mathrm{root}}$ across a sweep of reduced frequencies $k_G \in [0.3, 1.9]$, validated against Heathcote et al. \citep{validation}.}
    \label{fig:fig5}
\end{figure}

Figure \ref{fig:fig4_flex} plots the deflections of the flexible structure with $E = 3.4\times 10^9\text{ Pa}$. The flexible undergoes amplification in tip displacement, reaching peak $s/a_{\mathrm{root}} \approx 1.65$. Figure \ref{fig:fig4_inflex} plots the same for a less flexible plate with $E = 2.6\times 10^{10}\text{ Pa}$. Due to the flexural rigidity with an order of magnitude higher, the peak tip displacement remains small. The tip trajectory closely aligns with the prescribed root heaving kinematics, with ($s/a_{\mathrm{root}} \approx 1.05$). The solutions from \texttt{VarFlExI} align closely with the experimental measurements. A plausible cause for the small deviations in the phase of the subplots in Fig. \ref{fig:fig4} representing non-dimensional tip displacement with time might primarily be due to the ramping given to the heaving of the root. This introduces a phase offset between the root and the tip, especially in a water-tunnel setup, which will not align over time at steady state. Another cause may be the lack of effective modeling of the water medium's viscous effects. Despite minor physical simplifications, the agreement between the time-resolved waveforms and the reduced-frequency sweeps validates the partitioned \texttt{VarFlExI} framework for unsteady \textit{FSI} problems.

\subsection{Sensitivity Analysis}
\label{sec:convergence_analysis}

To establish the spatial mesh independence, wake fidelity, and temporal convergence of the proposed exchange interface, systematic sensitivity studies are conducted. The benchmark configuration models a flexible cantilever plate wing with the wing root rigidly clamped ($u_0 = v_0 = w = \theta_x = \theta_y = 0$ along $\eta^s = 0$) with zero prescribed root motion, and the structure is impulsively immersed in a uniform freestream flow. Unless otherwise stated, the sensitivity studies use the nominal configuration
listed in Table~\ref{tab:simulation_parameters}, with only the parameter under investigation varied. The structural plate domain is modeled as an isotropic Reissner-Mindlin continuum.

\begin{table*}[t]
\centering
\caption{Parameter ranges evaluated in the sensitivity studies. The baseline value used for each study is listed separately.}
\label{tab:sweep}

\renewcommand{\arraystretch}{1.2}

\begin{tabularx}{\textwidth}{l c X c}
\toprule
\textbf{Sensitivity Study} & \textbf{Symbol} &
\textbf{Swept Values} & \textbf{Baseline Value} \\
\midrule

Temporal Discretization
& $\Delta t$
& $\{0.001,\,0.010,\,0.050\}\text{ s}$
& $0.001\text{ s}$
\\

Fluid Surface Discretization
& $N_{\mathrm{span}}$
& $\{20,\,110,\,200\}$
& $80$
\\

Wake-Shedding Resolution
& $N_p^{(k)}-N_p^{(k-1)}$
& $\{1,\,2,\,3\}\text{ particles/step}$
& $1$
\\

\bottomrule
\end{tabularx}
\end{table*}

Unlike the heaving experiments in Section. (\ref{sec:validation}), wing root remains stationary, the structural response is driven entirely by the transient aerodynamic load buildup following the impulsive flow start. For consistent presentation across all sensitivity sweeps, physical time $t$ and transient wing-tip deflection $s_{\mathrm{tip}}(t)$ are non-dimensionalized using characteristic convective scales:

\begin{equation}
  t^* = \frac{t U_\infty}{c_{\mathrm{ref}}}, \qquad s^*_{\mathrm{tip}} = \frac{s_{\mathrm{tip}}}{c_{\mathrm{ref}}}.
  \label{eq:nondimensional_scales}
\end{equation}

\subsubsection{Temporal Discretization}
\label{sec:dt_convergence}

The temporal convergence of the partitioned solver is assessed using three primary coupling time steps, $\Delta t\in\{0.001,\,0.01,\,0.05\}\,\mathrm{s}$. An additional refinement with $\Delta t=0.0005\,\mathrm{s}$ is considered for the time-averaged spanwise loading analysis to provide an independent check of the selected
temporal resolution. Refer to Table (\ref{tab:sweep}) for the baseline values of the rest of the parameters used in this study.

\begin{figure}[pos=h!]
  \centering
  \includegraphics[width=0.95\linewidth]{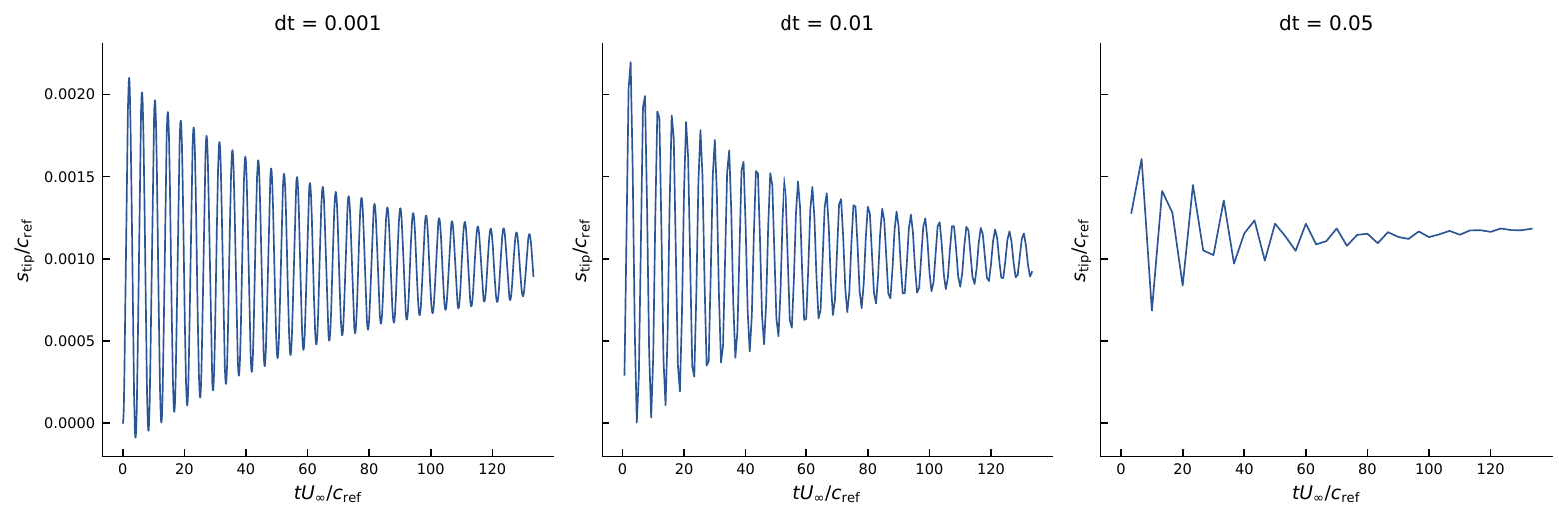}
  
  \caption{Normalized tip displacement $s_{\mathrm{tip}}/c_{\mathrm{ref}}$ as a function of
  non-dimensional convective time $tU_\infty/c_{\mathrm{ref}}$ for the three coupling time steps
  $\Delta t=0.050$, $0.010$, and $0.001\,\mathrm{s}$.}
  
  \label{fig:dt_sweep}
\end{figure}

\begin{figure}[pos=h!]
    \centering

    \begin{subfigure}[b]{0.48\textwidth}
        \centering
        \includegraphics[width=\textwidth, page=1]{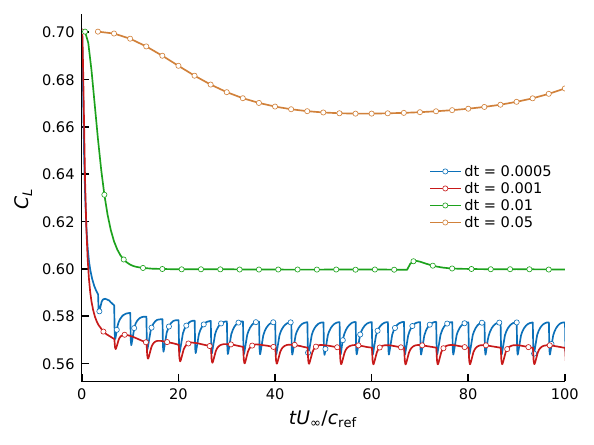}
        \caption{Temporal evolution of the total lift coefficient $C_L$ for the three coupling time steps considered in the transient convergence study.}
        \label{fig:dt_time}
    \end{subfigure}
    \hfill
    \begin{subfigure}[b]{0.48\textwidth}
        \centering
        \includegraphics[width=\textwidth, page=1]{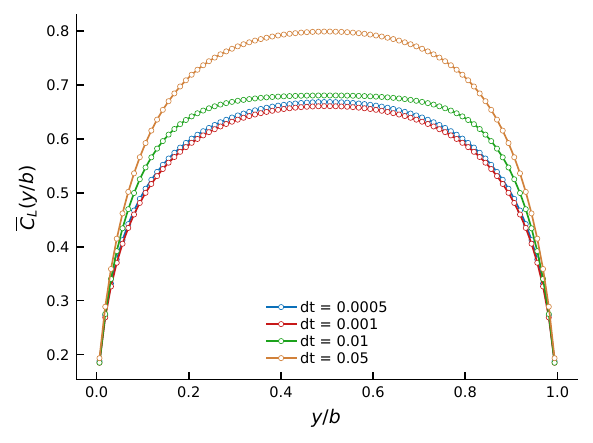}
        \caption{Time-averaged spanwise lift-coefficient $\overline{C}_L(y/b)$ distribution against spanwise stations.}
        \label{fig:dt_span}
    \end{subfigure}

    \caption{Temporal-discretization sensitivity of the aerodynamic response. The transient total
    lift coefficient is compared for the three time steps used in the coupled aeroelastic
    convergence study, while the spanwise loading distribution includes an additional finer
    time-step refinement to assess the convergence of the mean aerodynamic loading.}
    \label{fig:dt_cl}
\end{figure}

Figure~\ref{fig:dt_sweep} shows the transient non-dimensional tip displacement
$s_{\mathrm{tip}}/c_{\mathrm{ref}}$ over the convective time interval. Following the impulsive
start, the rapid development of the aerodynamic loading produces an initial increase in the tip
displacement, followed by decaying structural oscillations as the coupled system approaches a
quasi-steady aeroelastic state. The influence of the temporal resolution is most evident during
the initial transient, where the coarser time step produces a less resolved representation of the rapid structural response. Reducing the coupling time step from $\Delta t=0.01$ to $0.001\,\mathrm{s}$ produces a visibly
better-resolved transient and a smoother representation of the peak response. Further reduction
to $\Delta t=0.001\,\mathrm{s}$ produces only a small additional change in the predicted trajectory,
indicating that the principal features of the coupled structural response have become insensitive
to further refinement at this stage. Thus, $\Delta t=0.001\,\mathrm{s}$ is identified as the
finest time step required to resolve the structural transient within the three-case convergence
study.

To further verify the temporal resolution of the aerodynamic solution, the additional refinement with $\Delta t=0.0005\,\mathrm{s}$ is compared with the
$\Delta t=0.001\,\mathrm{s}$ case in Fig. (\ref{fig:dt_span}). The resulting time-averaged spanwise distribution is closely aligned with that obtained using $\Delta t=0.001\,\mathrm{s}$, indicating that further temporal refinement produces only a negligible change in the mean aerodynamic loading.
This additional refinement therefore provides an independent check that $\Delta t=0.001\,\mathrm{s}$ is sufficiently resolved for the subsequent coupled simulations. Based on the convergence of the transient structural response and the negligible change in the time-averaged spanwise aerodynamic loading upon refinement from $\Delta t=0.001$ to $0.0005\,\mathrm{s}$,
$\Delta t=0.001\,\mathrm{s}$ is selected as the baseline coupling time step for the subsequent simulations.


  
  

\subsubsection{Spatial Discretization}
\label{sec:dy_convergence}

The sensitivity of the aerodynamic solution to the spanwise discretization of the Actuator Model is evaluated using $N_{\mathrm{span}}\in\{20,\,110,\,200\}$ surface panels. The transient total lift coefficient and the time-averaged spanwise loading distribution are used to assess the effect of aerodynamic surface resolution.

\begin{figure}[pos=h!]
    \centering
    \begin{subfigure}[b]{0.48\textwidth}
        \centering
        \includegraphics[width=\textwidth, page=1]{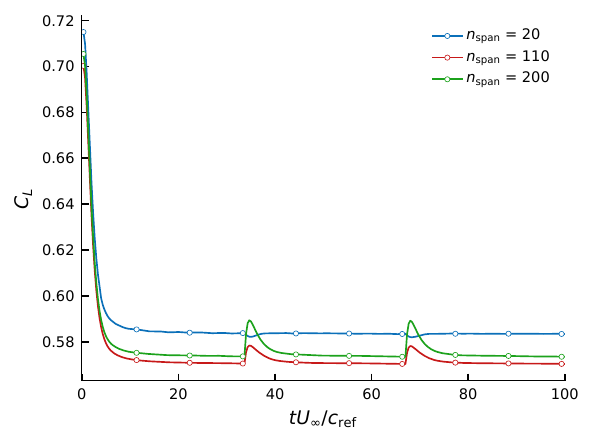}
        \caption{Temporal evolution of the total lift coefficient $C_L$ for the considered spanwise aerodynamic resolutions.}
        \label{fig:dy_time}
    \end{subfigure}
    \hfill
    \begin{subfigure}[b]{0.48\textwidth}
        \centering
        \includegraphics[width=\textwidth, page=1]{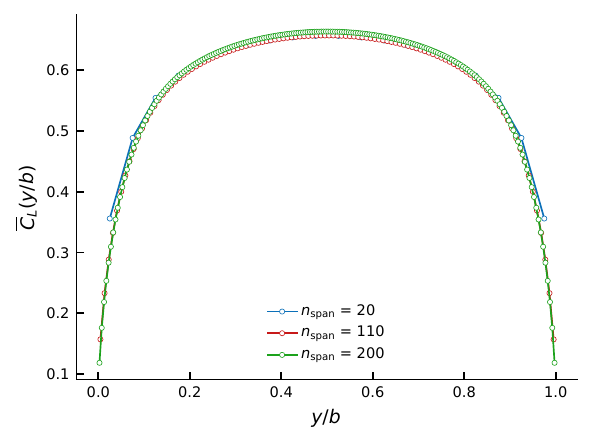}
        \caption{Time-averaged spanwise lift-coefficient $\overline{C}_L(y/b)$ distribution against spanwise stations.}
        \label{fig:dy_span}
    \end{subfigure}

    \caption{Sensitivity of the aerodynamic response to the spanwise discretization of the
    Actuator Surface Model for $N_{\mathrm{span}}=20$, $110$, and $200$ surface panels. The
    temporal evolution of the total lift coefficient and the corresponding time-averaged
    spanwise loading distribution are shown in (a) and (b), respectively.}
    \label{fig:dy_cl}
\end{figure}

Figure~\ref{fig:dy_cl}(a) shows the temporal evolution of the total lift coefficient for the
three spanwise resolutions. The coarsest discretization, $N_{\mathrm{span}}=20$, produces a
noticeably different aerodynamic response, particularly during the initial transient. Increasing
the resolution to $N_{\mathrm{span}}=110$ substantially reduces this difference, while further
refinement to $N_{\mathrm{span}}=200$ produces only a small additional change in the total lift
history. This indicates that the integrated aerodynamic response is largely insensitive to
further spanwise refinement beyond $N_{\mathrm{span}}=110$.

The corresponding time-averaged spanwise loading distributions are shown in
Fig.~\ref{fig:dy_cl}(b). The coarse $N_{\mathrm{span}}=20$ discretization exhibits larger differences in the resolved loading distribution, whereas the $N_{\mathrm{span}}=110$ and $N_{\mathrm{span}}=200$ results remain closely aligned over most of the span. The additional refinement from $110$ to $200$ panels produces only a minor change in both the coupled structural response and the mean aerodynamic loading.

The non-monotonic variation with increasing spanwise resolution is attributed to changes in the spatial sampling and quadrature of the non-uniform aerodynamic loading. Since the considered resolutions are not nested refinements, the resulting quadrature error need not decrease monotonically. Changes in the resolved circulation distribution and hence, wake shedding further affect the monotonicity. The close agreement between the $N_{\mathrm{span}}=110$ and $200$ cases, therefore, indicates practical resolution insensitivity rather than strict monotonic convergence.

\subsubsection{Wake Resolution}
\label{sec:pperstep_convergence}

The sensitivity of the coupled aerodynamic response to the wake-particle resolution is evaluated by varying the number of free vortex particles shed per time step. This study examines the effect of wake resolution on the aerodynamic loads while keeping the remaining numerical parameters unchanged. The corresponding structural tip-displacement histories were also examined for the three wake-shedding resolutions. The resulting trajectories were found to be nearly coincident over the simulated interval, indicating that the change in wake-particle resolution has only a minor effect on the global structural response for the present configuration. The aerodynamic coefficients are therefore considered more suitable for resolving the influence of wake resolution, and the temporal histories of the lift coefficient $C_L$ and the drag coefficient $C_D$, together with their corresponding time-averaged spanwise distributions, are used for the convergence assessment.

\begin{figure}[pos=h!]
    \centering

    \begin{subfigure}[b]{0.48\textwidth}
        \centering
        \includegraphics[width=\textwidth]{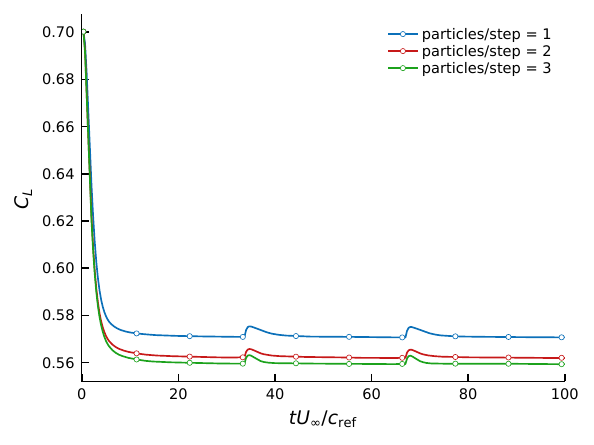}
        \caption{Temporal evolution of the total lift coefficient $C_L$ for the considered
        wake-shedding resolutions.}
        \label{fig:pperstep_cl_time}
    \end{subfigure}
    \hfill
    \begin{subfigure}[b]{0.48\textwidth}
        \centering
        \includegraphics[width=\textwidth]{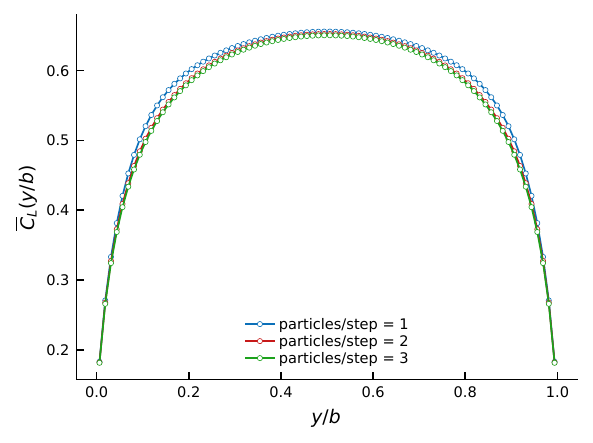}
        \caption{Time-averaged spanwise lift-coefficient $\overline{C}_L(y/b)$ distribution against spanwise stations.}
        \label{fig:pperstep_cl_span}
    \end{subfigure}

    \vspace{4mm}

    \begin{subfigure}[b]{0.48\textwidth}
        \centering
        \includegraphics[width=\textwidth]{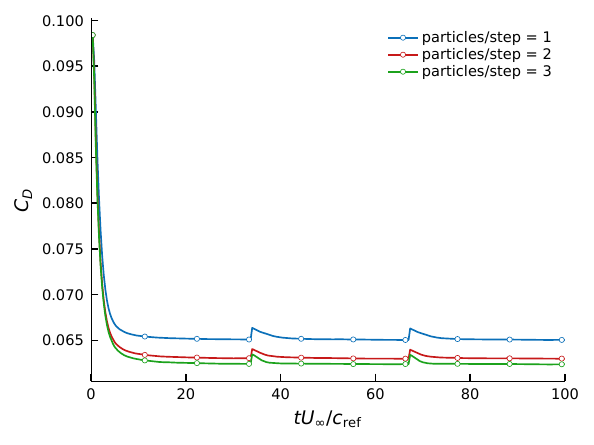}
        \caption{Temporal evolution of the total drag coefficient $C_D$ for the considered
        wake-shedding resolutions.}
        \label{fig:pperstep_cd_time}
    \end{subfigure}
    \hfill
    \begin{subfigure}[b]{0.48\textwidth}
        \centering
        \includegraphics[width=\textwidth]{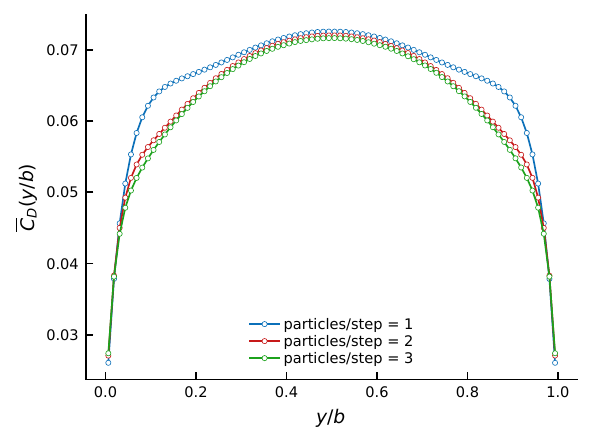}
        \caption{Time-averaged spanwise lift-coefficient $\overline{C}_D(y/b)$ distribution against spanwise stations.}
        \label{fig:pperstep_cd_span}
    \end{subfigure}

    \caption{Sensitivity of the aerodynamic response to the wake-shedding resolution.
    The total lift and drag histories and their corresponding time-averaged spanwise
    distributions are compared for one, two, and three free vortex particles shed per time step.}
    \label{fig:pperstep}
\end{figure}

Figure~\ref{fig:pperstep} shows the influence of wake-particle resolution on the aerodynamic response. The temporal histories of $C_L$ and $C_D$ exhibit the largest differences for the coarsest wake representation, while the solutions obtained with two and three particles shed per time step remain closely aligned. The corresponding time-averaged spanwise distributions show the same trend, with the loading distributions becoming progressively less sensitive to further increase in the wake-particle resolution. The close agreement between the two and three-particle cases indicates that increasing the wake resolution beyond two particles per time step produces only a limited change in the predicted aerodynamic response for the present configuration. Consequently, to balance accuracy and computational cost, we choose either 1 or 2 particles to shed per timestep and per panel for all simulation runs.


\section{Conclusions}
\label{sec:conc}

In this work, we present $\texttt{VarFlExI}$, an explicit staggered coupled aeroelastic framework for modeling aeroelastic flutter, by implementing an exchange interface between open source frameworks, \texttt{FLOWUnsteady} and \texttt{FEniCS}, as aerodynamic and structural solvers, respectively. Our aerodynamic and structural solvers are coupled through an explicit staggered partitioned scheme, ensuring conservation of virtual work for load and displacement transfer across the non-matching interfaces through a Common Refinement Operator. We successfully validate our framework against water-tunnel experiments, demonstrating accurate prediction of the coupled aeroelastic response. The computational efficiency of our meshless aerodynamic solver enables simulations at significantly lower computational cost while maintaining accuracy, whereas the 2-D structural solver significantly lowers computational costs compared to high fidelty 3-D discretizations, while demonstrating comparable structural response.

\paragraph{Future Work:} We intend to parallelize the \texttt{VarFlExI} framework to speed up computationally heavy algorithms in each vertical. Prior to that, we have to identify the computational bottlenecks through time complexity checks by running systematic parameter sweeps, primarily on particles shed per step and discretizations. A crucial modeling aspect remaining is to introduce wing-aware geometry in the structural solver, replacing the existing mid-plane 2D approximation.

\paragraph{Limitations:} The present structural formulation is restricted to plate-like, predominantly two-dimensional structural behavior and does not resolve fully three-dimensional through-thickness deformation or localized structural phenomena. Its applicability is consequently restricted to deformation regimes in which the underlying first-order shear-deformation assumptions remain appropriate; strongly three-dimensional or highly localized structural responses would require higher-order plate theories or a full three-dimensional continuum formulation.

The present actuator-based \textit{rVPM} formulation is intended primarily for incompressible, vortex-dominated unsteady flows with predominantly attached aerodynamic loading. The actuator representation does not explicitly resolve boundary-layer development or the onset of flow
separation and therefore cannot reproduce the detailed formation and evolution of separation wakes, including leading-edge separation vortices. Sectional viscous and parasitic effects are instead incorporated through prescribed two-dimensional airfoil polar lookup tables. The optional actuator-surface formulation in our framework provides a more distributed representation of bound vorticity when required.


\section{Acknowledgments}
The structural solver is implemented using the \texttt{FEniCS} Project \citep{ufl,automated_solution}, an open-source computing platform for solving partial differential equations using the finite element method. It is integrated with the open-source \texttt{ParaView} \citep{paraview} as its visualization engine. \texttt{FEniCS} provides a high-level interface for defining variational problems through the Unified Form Language (\citep{ufl}), enabling concise and mathematically consistent implementation of continuum mechanics models. In the present work, \texttt{FEniCS} along with the \texttt{fenics\_shells} library \citep{fenics_article}, is used to construct a three-dimensional elastodynamic solver based on a continuous Galerkin finite element formulation, with the elastodynamics integration solved with $\texttt{MUMPS}$ (MUltifrontal Massively Parallel sparse direct Solver).

The fluid solver is implemented using \texttt{FLOWUnsteady} \citep{alvarez2022flowunsteady, FLOWUnsteady}, an open-source aerodynamic simulation framework built in the \texttt{Julia} programming language \citep{julia}. \texttt{FLOWUnsteady} provides a high-level, modular architecture for simulating complex unsteady, turbulent flows without body-fitted volume meshes. In the present work, \texttt{FLOWUnsteady}, along with the underlying \texttt{FLOWVLM} library for lifting-surface kinematics and the \texttt{vpm} module for Fast Multipole Method (FMM) particle interactions, is used to construct a three-dimensional fluid solver based on the reformulated Vortex Particle Method (\textit{rVPM}) and Actuator Surface Modeling.

\bibliographystyle{cas-model2-names}
\bibliography{references}

@article{chai_flutter,
author = {Chai, Yuyang and Gao, Wei and Ankay, Benjamin and Li, Fengming and Zhang, Chuanzeng},
title = {Aeroelastic analysis and flutter control of wings and panels: A review},
journal = {International Journal of Mechanical System Dynamics},
volume = {1},
number = {1},
pages = {5-34},
doi = {https://doi.org/10.1002/msd2.12015},
url = {https://onlinelibrary.wiley.com/doi/abs/10.1002/msd2.12015},
eprint = {https://onlinelibrary.wiley.com/doi/pdf/10.1002/msd2.12015},
year = {2021}
}

@article{raveh_intro,
title = {Computational-fluid-dynamics-based aeroelastic analysis and structural design optimization—a researcher’s perspective},
journal = {Computer Methods in Applied Mechanics and Engineering},
volume = {194},
number = {30},
pages = {3453-3471},
year = {2005},
note = {Structural and Design Optimization},
issn = {0045-7825},
doi = {https://doi.org/10.1016/j.cma.2004.12.027},
url = {https://www.sciencedirect.com/science/article/pii/S0045782505000563},
author = {Daniella E. Raveh},
}

@article{kamakoti_intro,
title = {Fluid–structure interaction for aeroelastic applications},
journal = {Progress in Aerospace Sciences},
volume = {40},
number = {8},
pages = {535-558},
year = {2004},
issn = {0376-0421},
doi = {https://doi.org/10.1016/j.paerosci.2005.01.001},
url = {https://www.sciencedirect.com/science/article/pii/S0376042105000084},
author = {Ramji Kamakoti and Wei Shyy},
}

@article{higharwing,
title = {A review on non-linear aeroelasticity of high aspect-ratio wings},
journal = {Progress in Aerospace Sciences},
volume = {89},
pages = {40-57},
year = {2017},
issn = {0376-0421},
doi = {https://doi.org/10.1016/j.paerosci.2016.12.004},
url = {https://www.sciencedirect.com/science/article/pii/S037604211630077X},
author = {Frederico Afonso and José Vale and Éder Oliveira and Fernando Lau and Afzal Suleman},
}

@article{rom_intro1,
title = {Efficient aeroelastic reduced order model with global structural modifications},
journal = {Aerospace Science and Technology},
volume = {76},
pages = {1-13},
year = {2018},
issn = {1270-9638},
doi = {https://doi.org/10.1016/j.ast.2018.01.023},
url = {https://www.sciencedirect.com/science/article/pii/S1270963817308003},
author = {Gang Chen and Dongfeng Li and Qiang Zhou and Andrea {Da Ronch} and Yueming Li},
}

@article{rom_intro2,
title = {Aeroelastic reduced-order modeling for efficient static aeroelastic analysis considering geometric nonlinearity},
journal = {Journal of Fluids and Structures},
volume = {124},
pages = {104055},
year = {2024},
issn = {0889-9746},
doi = {https://doi.org/10.1016/j.jfluidstructs.2023.104055},
url = {https://www.sciencedirect.com/science/article/pii/S0889974623002232},
author = {Kai Li and Jiaqing Kou and Weiwei Zhang},
}

@article{fmm,
    author = {J barnes},
    title = {A hierarchical O(N log N) force-calculation algorithm},
    journal = {Nature 324},
    year = {1986}
}

@inproceedings{AndrewFLOWFMM,
    author = {Ryan Anderson and Andrew Ning},
    title = {Solving Unsteady Potential Flow Problems in O(n) Time},
    booktitle = {AIAA Aviation Forum},
    year = {2024}
}

@article{control_aero_brunton,
  author  = {Hickner, Michelle and Fasel, Urban and Nair, Aditya G. and Brunton, Bingni W. and Brunton, Steven L.},
  title   = {Data-Driven Unsteady Aeroelastic Modeling for Control},
  journal = {AIAA Journal},
  volume  = {61},
  number  = {2},
  pages   = {780--792},
  year    = {2023},
  doi     = {10.2514/1.J061518},
  url     = {https://doi.org/10.2514/1.J061518}
}

@article{jax_aero_intro,
title = {JAX-based aeroelastic simulation engine for differentiable aircraft dynamics},
journal = {Computer Physics Communications},
volume = {311},
pages = {109547},
year = {2025},
issn = {0010-4655},
doi = {https://doi.org/10.1016/j.cpc.2025.109547},
url = {https://www.sciencedirect.com/science/article/pii/S0010465525000505},
author = {Alvaro Cea and Rafael Palacios},
}

@article{shape_opt,
title = {Efficient aerodynamic shape optimization in MDO context},
journal = {Journal of Computational and Applied Mathematics},
volume = {203},
number = {2},
pages = {548-560},
year = {2007},
note = {Special Issue: The first Indo-German Conference on PDE, Scientific Computing and Optimization in Applications},
issn = {0377-0427},
doi = {https://doi.org/10.1016/j.cam.2006.04.013},
url = {https://www.sciencedirect.com/science/article/pii/S0377042706002214},
author = {Antonio Fazzolari and Nicolas R. Gauger and Joël Brezillon},
}

@article{vpm_intro_wind,
title = {Pseudo three-dimensional simulation of aeroelastic response to turbulent wind using Vortex Particle Methods},
journal = {Journal of Fluids and Structures},
volume = {72},
pages = {1-24},
year = {2017},
issn = {0889-9746},
doi = {https://doi.org/10.1016/j.jfluidstructs.2017.04.001},
url = {https://www.sciencedirect.com/science/article/pii/S0889974616303899},
author = {Khaled {Ibrahim Tolba} and Guido Morgenthal},
}

@article{chau_harvester_vpm,
author = {Samir Chawdhury and Guido Morgenthal},
title ={Numerical simulations of aeroelastic instabilities to optimize the performance of flutter-based electromagnetic energy harvesters},

journal = {Journal of Intelligent Material Systems and Structures},
volume = {29},
number = {4},
pages = {479-495},
year = {2018},
doi = {10.1177/1045389X17711784},

URL = {   
        https://doi.org/10.1177/1045389X17711784
},
eprint = { 
        https://doi.org/10.1177/1045389X17711784
}
,
}

@article{vpm_appln_intro2,
title = {Mid-fidelity numerical approach for the investigation of wing-propeller aerodynamic interaction},
journal = {Aerospace Science and Technology},
volume = {146},
pages = {108950},
year = {2024},
issn = {1270-9638},
doi = {https://doi.org/10.1016/j.ast.2024.108950},
url = {https://www.sciencedirect.com/science/article/pii/S127096382400083X},
author = {Claudio Niro and Alberto Savino and Alessandro Cocco and Alex Zanotti},
}

@phdthesis{FLOWUnsteady,
    author = {Eduardo J. Alvarez},
    title = {Reformulated Vortex Particle Method and Meshless Large Eddy Reformulated Vortex Particle Method and Meshless Large Eddy Simulation of Multirotor Aircraft },
    school = {Brigham Young University },
    year = {2022}
}

@inproceedings{alvarez2022flowunsteady,
  author    = {Alvarez, Eduardo J. and Mehr, Judd and Ning, Andrew},
  title     = {{FLOWUnsteady}: An Interactional Aerodynamics Solver for Multirotor Aircraft and Wind Energy},
  booktitle = {AIAA AVIATION 2022 Forum},
  address   = {Chicago, IL},
  month     = {June},
  year      = {2022},
  doi       = {10.2514/6.2022-3218},
  url       = {https://doi.org}
}

@inbook{aeroelasticmdo,
    author = {Sebastiaan P. van Schie and Han Zhao and Jiayao Yan and Ru Xiang and John T. Hwang and David Kamensky},
    title = {Solver-Independent Aeroelastic Coupling For Large-Scale Multidisciplinary Design Optimization},
    booktitle = {AIAA SCITECH 2023 Forum},
    doi = {10.2514/6.2023-0727},
    URL = {https://arc.aiaa.org/doi/abs/10.2514/6.2023-0727},
    eprint = {https://arc.aiaa.org/doi/pdf/10.2514/6.2023-0727}
}

@book{Katz_Plotkin_2001, place={Cambridge}, edition={2}, series={Cambridge Aerospace Series}, title={Low-Speed Aerodynamics}, publisher={Cambridge University Press}, author={Katz, Joseph and Plotkin, Allen}, year={2001}, collection={Cambridge Aerospace Series}}

@inbook{vortex_winckelmans,
author = {Winckelmans, G. S.},
publisher = {John Wiley \& Sons, Ltd},
isbn = {9780470091357},
title = {Vortex Methods},
booktitle = {Encyclopedia of Computational Mechanics},
chapter = {5},
pages = {},
doi = {https://doi.org/10.1002/0470091355.ecm055},
url = {https://onlinelibrary.wiley.com/doi/abs/10.1002/0470091355.ecm055},
eprint = {https://onlinelibrary.wiley.com/doi/pdf/10.1002/0470091355.ecm055},
year = {2004},
}

@article{leonard_vortex,
title = {Vortex methods for flow simulation},
journal = {Journal of Computational Physics},
volume = {37},
number = {3},
pages = {289-335},
year = {1980},
issn = {0021-9991},
doi = {https://doi.org/10.1016/0021-9991(80)90040-6},
url = {https://www.sciencedirect.com/science/article/pii/0021999180900406},
author = {A Leonard},
}

@article{mansfield_vortex,
title = {A Dynamic LES Scheme for the Vorticity Transport Equation: Formulation anda PrioriTests},
journal = {Journal of Computational Physics},
volume = {145},
number = {2},
pages = {693-730},
year = {1998},
issn = {0021-9991},
doi = {https://doi.org/10.1006/jcph.1998.6051},
url = {https://www.sciencedirect.com/science/article/pii/S002199919896051X},
author = {John R. Mansfield and Omar M. Knio and Charles Meneveau},
}

@article{winckelmans,
author = {Winckelmans, G. S. and Leonard, A.},
title = {Contributions to vortex particle methods for the computation of three-dimensional incompressible unsteady flows},
year = {1993},
issue_date = {Dec. 1993},
publisher = {Academic Press Professional, Inc.},
address = {USA},
volume = {109},
number = {2},
issn = {0021-9991},
url = {https://doi.org/10.1006/jcph.1993.1216},
doi = {10.1006/jcph.1993.1216},
journal = {J. Comput. Phys.},
month = dec,
pages = {247–273},
numpages = {27}
}

@inproceedings{ASM,
    author = {Shen, W. Z. and S{\o}rensen, J. N. and Zhang, J.},
    title = {Actuator surface model for wind turbine flow computations},
    booktitle = {Proceedings of European Wind Energy Conference 2007},
    year = {2007}
}

@article{validation,
title = {Effect of spanwise flexibility on flapping wing propulsion},
journal = {Journal of Fluids and Structures},
volume = {24},
number = {2},
pages = {183-199},
year = {2008},
issn = {0889-9746},
doi = {https://doi.org/10.1016/j.jfluidstructs.2007.08.003},
url = {https://www.sciencedirect.com/science/article/pii/S0889974607000692},
author = {S. Heathcote and Z. Wang and I. Gursul}
}

@book{pistolesi,
    author = {E. Pistolesi},
    title = {Ground effect-theory and practice},
    publisher = {},
    year = {1937}
}

@article{3dviscous_nonlinearpanel,
  title={Development of a three-dimensional viscous aeroelastic solver for nonlinear panel flutter},
  author={Gordnier, Raymond E and Visbal, Miguel R},
  journal={Journal of fluids and structures},
  volume={16},
  number={4},
  pages={497--527},
  year={2002},
  publisher={Elsevier}
}

@article{farhat_workcons,
title = {Load and motion transfer algorithms for fluid/structure interaction problems with non-matching discrete interfaces: Momentum and energy conservation, optimal discretization and application to aeroelasticity},
journal = {Computer Methods in Applied Mechanics and Engineering},
volume = {157},
number = {1},
pages = {95-114},
year = {1998},
issn = {0045-7825},
doi = {https://doi.org/10.1016/S0045-7825(97)00216-8},
url = {https://www.sciencedirect.com/science/article/pii/S0045782597002168},
author = {C. Farhat and M. Lesoinne and P. {Le Tallec}},
}

@article{rbf_workcons,
author = {Costin, W. J. and Allen, C. B.},
title = {Numerical study of radial basis function interpolation for data transfer across discontinuous mesh interfaces},
journal = {International Journal for Numerical Methods in Fluids},
volume = {72},
number = {10},
pages = {1076-1095},
doi = {https://doi.org/10.1002/fld.3778},
url = {https://onlinelibrary.wiley.com/doi/abs/10.1002/fld.3778},
eprint = {https://onlinelibrary.wiley.com/doi/pdf/10.1002/fld.3778},
year = {2013}
}

@article{flyer_rbf,
title = {On the role of polynomials in RBF-FD approximations: I. Interpolation and accuracy},
journal = {Journal of Computational Physics},
volume = {321},
pages = {21-38},
year = {2016},
issn = {0021-9991},
doi = {https://doi.org/10.1016/j.jcp.2016.05.026},
url = {https://www.sciencedirect.com/science/article/pii/S0021999116301632},
author = {Natasha Flyer and Bengt Fornberg and Victor Bayona and Gregory A. Barnett},
}

@article{ufl,
  title={Unified form language: A domain-specific language for weak formulations of partial differential equations},
  author={Aln{\ae}s, Martin S and Logg, Anders and {\O}lgaard, Kristian B and Rognes, Marie E and Wells, Garth N},
  journal={ACM Transactions on Mathematical Software (TOMS)},
  volume={40},
  number={2},
  pages={1--37},
  year={2014},
  publisher={ACM New York, NY, USA}
}

@article{fenics,
  title     = {The {FEniCS} Project Version 1.5},
  author    = {Alnaes, Martin S. and Blechta, Jan and Hake, Johan and Johansson, August and Kehlet, Benjamin and Logg, Anders and Richardson, Chris N. and Ring, Johannes and Rognes, Marie E. and Wells, Garth N.},
  journal   = {Archive of Numerical Software},
  year      = {2015},
  volume    = {3},
  doi       = {10.11588/ans.2015.100.20553},
}

@book{automated_solution,
  title     = {Automated Solution of Differential Equations by the Finite Element Method},
  author    = {Logg, Anders and Mardal, Kent-Andre and Wells, Garth N. and others},
  year      = {2012},
  doi       = {10.1007/978-3-642-23099-8},
  publisher = {Springer},
}

@article{gen-alpha,
      author = {Erlicher, Silvano and Bonaventura, Luca and Bursi, Omar S.},
      title = {The analysis of the Generalized-$\alpha$ method for non-linear dynamic problems},
      journal = {Computational Mechanics},
      volume = {28},
      number = {2},
      pages = {83--104},
      year = {2002},
      doi = {10.1007/S00466-001-0273-Z}
}

@misc{fenics_article,
    title = {{FEniCS}-{Shells}},
    url = {https://figshare.com/articles/FEniCS-Shells/4291160},
    author = {Hale, Jack S. and Brunetti, Matteo and Bordas, Stéphane P.A. and Maurini, Corrado},
    month = dec,
    year = {2016},
    doi = {10.6084/m9.figshare.4291160}
}

@article{structural_dynamics,
    author = {Nathan M. Newmark },
    title = {A Method of Computation for Structural Dynamics},
    journal = {Journal of the Engineering Mechanics Division},
    volume = {85},
    number = {3},
    pages = {67-94},
    year = {1959},
    doi = {10.1061/JMCEA3.0000098},
    URL = {https://ascelibrary.com/doi/abs/10.1061/JMCEA3.0000098},
    eprint = {https://ascelibrary.com/doi/pdf/10.1061/JMCEA3.0000098}
}

@article{unsteadyvlm,
    title = {Applications of the unsteady vortex-lattice method in aircraft aeroelasticity and flight dynamics},
    journal = {Progress in Aerospace Sciences},
    volume = {55},
    pages = {46-72},
    year = {2012},
    issn = {0376-0421},
    doi = {https://doi.org/10.1016/j.paerosci.2012.06.001},
    url = {https://www.sciencedirect.com/science/article/pii/S0376042112000620},
    author = {Joseba Murua and Rafael Palacios and J. Michael R. Graham}
}

@article{crm,
title = {A 3D common-refinement method for non-matching meshes in partitioned variational fluid–structure analysis},
journal = {Journal of Computational Physics},
volume = {374},
pages = {163-187},
year = {2018},
issn = {0021-9991},
doi = {https://doi.org/10.1016/j.jcp.2018.05.023},
url = {https://www.sciencedirect.com/science/article/pii/S0021999118303231},
author = {Yulong Li and Yun Zhi Law and Vaibhav Joshi and Rajeev K. Jaiman},
}

@article{vortex_methods,
doi = {10.1088/0957-0233/12/3/704},
url = {https://doi.org/10.1088/0957-0233/12/3/704},
year = {2001},
month = {mar},
publisher = {},
volume = {12},
number = {3},
pages = {354},
author = {G-H Cottet and P D Koumoutsakos},
title = {Vortex Methods: Theory and Practice},
journal = {Measurement Science and Technology},
}

@article{nguyen2021stable,
  title={Stable and accurate numerical methods for generalized Kirchhoff--Love plates: DTA Nguyen et al.},
  author={Nguyen, Duong TA and Li, Longfei and Ji, Hangjie},
  journal={Journal of Engineering Mathematics},
  volume={130},
  number={1},
  pages={6},
  year={2021},
  publisher={Springer}
}

@article{GORDNIER2002497,
title = {DEVELOPMENT OF A THREE-DIMENSIONAL VISCOUS AEROELASTIC SOLVER FOR NONLINEAR PANEL FLUTTER},
journal = {Journal of Fluids and Structures},
volume = {16},
number = {4},
pages = {497-527},
year = {2002},
issn = {0889-9746},
doi = {https://doi.org/10.1006/jfls.2000.0434},
url = {https://www.sciencedirect.com/science/article/pii/S0889974600904341},
author = {R.E. GORDNIER and M.R. VISBAL},
}

@article{article_reisner,
author = {Lyly, Mikko and Stenberg, Rolf and Vihinen, Teemu},
year = {1993},
month = {12},
pages = {343-357},
title = {A stable bilinear element for Reissner–Mindlin plate model},
volume = {110},
journal = {Computer Methods in Applied Mechanics and Engineering - COMPUT METHOD APPL MECH ENG},
doi = {10.1016/0045-7825(93)90214-I}
}

@article{article_reisner_kirchoff,
author = {Arnold, Douglas and Madureira, Alexandre and Zhang, Sheng},
year = {2002},
month = {06},
pages = {171-185},
title = {On the Range of Applicability of the Reissner–Mindlin and Kirchhoff–Love Plate Bending Models},
volume = {67},
journal = {Journal of Elasticity (2002)},
doi = {10.1023/A:1024986427134}
}

@InProceedings{formulation_thickthin,
author="Chen, D. P.
and Pan, Y. S.",
editor="Yagawa, Genki
and Atluri, Satya N.",
title="Formulation of Reissner-Mindlin Moderately-Thick/Thin Plate Bending Elements",
booktitle="Computational Mechanics '86",
year="1986",
publisher="Springer Japan",
address="Tokyo",
pages="49--54",
isbn="978-4-431-68042-0"
}

@article{influence_rotatory,
    author = {Mindlin, R. D.},
    title = {Influence of Rotatory Inertia and Shear on Flexural Motions of Isotropic, Elastic Plates},
    journal = {Journal of Applied Mechanics},
    volume = {18},
    number = {1},
    pages = {31-38},
    year = {2021},
    month = {04},
    issn = {0021-8936},
    doi = {10.1115/1.4010217},
    url = {https://doi.org/10.1115/1.4010217},
}

@article{rayleigh,
author = {Alipour, A and Zareian, Farzin and Student, Phd},
year = {2008},
month = {01},
pages = {},
title = {Study Rayleigh damping in structures; unceratinties and treatments}
}

@article{julia,
author = {Bezanson, Jeff and Edelman, Alan and Karpinski, Stefan and Shah, Viral B.},
title = {Julia: A Fresh Approach to Numerical Computing},
journal = {SIAM Review},
volume = {59},
number = {1},
pages = {65-98},
year = {2017},
doi = {10.1137/141000671},

URL = { 
    
        https://doi.org/10.1137/141000671
    
    

},
eprint = { 
    
        https://doi.org/10.1137/141000671
    
    

}
}

@article{li2019novel,
  title={A novel 3D variational aeroelastic framework for flexible multibody dynamics: Application to bat-like flapping dynamics},
  author={Li, Guojun and Law, Yun Zhi and Jaiman, Rajeev K},
  journal={Computers \& Fluids},
  volume={180},
  pages={96--116},
  year={2019},
  publisher={Elsevier}
}

@article{paraview,
author = {Ahrens, J. and Geveci, Berk and Law, Charles},
year = {2005},
month = {01},
pages = {},
title = {ParaView: An End-User Tool for Large Data Visualization},
journal = {Visualization Handbook}
}

@article{gen_alpha,
author = {Su, Weihua and Cesnik, Carlos},
year = {2011},
month = {08},
pages = {2349-2360},
title = {Strain-Based Geometrically Nonlinear Beam Formulation for Modeling Very Flexible Aircraft},
volume = {48},
journal = {International Journal of Solids and Structures},
doi = {10.1016/j.ijsolstr.2011.04.012}
}

@article{SLATTERY2016164,
title = {Mesh-free data transfer algorithms for partitioned multiphysics problems: Conservation, accuracy, and parallelism},
journal = {Journal of Computational Physics},
volume = {307},
pages = {164-188},
year = {2016},
issn = {0021-9991},
doi = {https://doi.org/10.1016/j.jcp.2015.11.055},
url = {https://www.sciencedirect.com/science/article/pii/S0021999115008037},
author = {Stuart R. Slattery}
}

@article{SHAFAGHAT2022107663,
title = {Nonlinear aeroelastic analysis of a HALE aircraft with flexible components},
journal = {Aerospace Science and Technology},
volume = {127},
pages = {107663},
year = {2022},
issn = {1270-9638},
doi = {https://doi.org/10.1016/j.ast.2022.107663},
url = {https://www.sciencedirect.com/science/article/pii/S1270963822003376},
author = {S. Shafaghat and M.A. Noorian and S. Irani}
}

@article{VINDIGNI2025107618,
title = {A refined aeroelastic beam finite element for the stability analysis of flexible subsonic wings},
journal = {Computers \& Structures},
volume = {307},
pages = {107618},
year = {2025},
issn = {0045-7949},
doi = {https://doi.org/10.1016/j.compstruc.2024.107618},
url = {https://www.sciencedirect.com/science/article/pii/S004579492400347X},
author = {Carmelo Rosario Vindigni and Giuseppe Mantegna and Calogero Orlando and Andrea Alaimo and Marco Berci},
}

@article{KAMPCHEN200363,
title = {Dynamic aero-structural response of an elastic wing model},
journal = {Journal of Fluids and Structures},
volume = {18},
number = {1},
pages = {63-77},
year = {2003},
issn = {0889-9746},
doi = {https://doi.org/10.1016/S0889-9746(03)00090-2},
url = {https://www.sciencedirect.com/science/article/pii/S0889974603000902},
author = {M. Kämpchen and A. Dafnis and H.-G. Reimerdes and G. Britten and J. Ballmann}
}

@article{GILLEBAART2016512,
title = {Low-fidelity 2D isogeometric aeroelastic analysis and optimization method with application to a morphing airfoil},
journal = {Computer Methods in Applied Mechanics and Engineering},
volume = {305},
pages = {512-536},
year = {2016},
issn = {0045-7825},
doi = {https://doi.org/10.1016/j.cma.2016.03.014},
url = {https://www.sciencedirect.com/science/article/pii/S0045782516300962},
author = {E. Gillebaart and R. {De Breuker}}
}

@article{RAGAB2019299,
title = {Finite element analysis of aero-hydroelastic stability of arbitrary shape panels},
journal = {Aerospace Science and Technology},
volume = {90},
pages = {299-313},
year = {2019},
issn = {1270-9638},
doi = {https://doi.org/10.1016/j.ast.2019.04.033},
url = {https://www.sciencedirect.com/science/article/pii/S1270963818326208},
author = {Saad A. Ragab and Hassan E. Fayed}
}

@article{LI2024104055,
title = {Aeroelastic reduced-order modeling for efficient static aeroelastic analysis considering geometric nonlinearity},
journal = {Journal of Fluids and Structures},
volume = {124},
pages = {104055},
year = {2024},
issn = {0889-9746},
doi = {https://doi.org/10.1016/j.jfluidstructs.2023.104055},
url = {https://www.sciencedirect.com/science/article/pii/S0889974623002232},
author = {Kai Li and Jiaqing Kou and Weiwei Zhang}
}

@article{PEREIRA2025113185,
title = {Modified consistent element-free Galerkin method applied to Reissner–Mindlin plates},
journal = {Thin-Walled Structures},
volume = {212},
pages = {113185},
year = {2025},
issn = {0263-8231},
doi = {https://doi.org/10.1016/j.tws.2025.113185},
url = {https://www.sciencedirect.com/science/article/pii/S0263823125002794},
author = {Marcelo Silveira Pereira and Mauricio Vicente Donadon}
}

@article{crm_initial,
author = {Jaiman, R. K. and Jiao, X. and Geubelle, P. H. and Loth, E.},
title = {Assessment of conservative load transfer for fluid–solid interface with non-matching meshes},
journal = {International Journal for Numerical Methods in Engineering},
volume = {64},
number = {15},
pages = {2014-2038},
doi = {https://doi.org/10.1002/nme.1434},
url = {https://onlinelibrary.wiley.com/doi/abs/10.1002/nme.1434},
eprint = {https://onlinelibrary.wiley.com/doi/pdf/10.1002/nme.1434},
year = {2005}
}

@article{DEBOER20084284,
title = {Comparison of conservative and consistent approaches for the coupling of non-matching meshes},
journal = {Computer Methods in Applied Mechanics and Engineering},
volume = {197},
number = {49},
pages = {4284-4297},
year = {2008},
issn = {0045-7825},
doi = {https://doi.org/10.1016/j.cma.2008.05.001},
url = {https://www.sciencedirect.com/science/article/pii/S0045782508001916},
author = {A. {de Boer} and A.H. {van Zuijlen} and H. Bijl}
}

@article{LOMBARDI2013117,
title = {Radial basis functions for inter-grid interpolation and mesh motion in FSI problems},
journal = {Computer Methods in Applied Mechanics and Engineering},
volume = {256},
pages = {117-131},
year = {2013},
issn = {0045-7825},
doi = {https://doi.org/10.1016/j.cma.2012.12.019},
url = {https://www.sciencedirect.com/science/article/pii/S0045782513000029},
author = {M. Lombardi and N. Parolini and A. Quarteroni}
}

@article{LIU2017810,
title = {Time efficient aeroelastic simulations based on radial basis functions},
journal = {Journal of Computational Physics},
volume = {330},
pages = {810-827},
year = {2017},
issn = {0021-9991},
doi = {https://doi.org/10.1016/j.jcp.2016.10.063},
url = {https://www.sciencedirect.com/science/article/pii/S002199911630571X},
author = {Wen Liu and ChengDe Huang and Guowei Yang}
}

@article{PARENTEAU2020103146,
title = {A general modal frequency-domain vortex lattice method for aeroelastic analyses},
journal = {Journal of Fluids and Structures},
volume = {99},
pages = {103146},
year = {2020},
issn = {0889-9746},
doi = {https://doi.org/10.1016/j.jfluidstructs.2020.103146},
url = {https://www.sciencedirect.com/science/article/pii/S0889974620306150},
author = {Matthieu Parenteau and Eric Laurendeau}
}

\appendix
\section{Appendix}

\subsection{Viscous Diffusion}
\label{app:viscous}

Viscous diffusion in the \textit{rVPM} is treated using the core-spreading method, in which viscous diffusion is represented through the evolution of the particle core size \citep{winckelmans,alvarez2022flowunsteady}.
For the Gaussian kernel adopted in \texttt{FLOWUnsteady}, the viscous contribution to the core-radius evolution is,

\begin{equation}
  \left(\frac{\mathrm{d}\sigma_p}{\mathrm{d}t}\right)_{\mathrm{viscous}}
  = \frac{2\nu_f}{\sigma_p}.
  \label{eq:core_spreading_rate}
\end{equation}

The viscous contribution is combined with the inviscid evolution of $\sigma_p$ during time integration. The core-spreading scheme is coupled with radial-basis-function interpolation for meshless spatial adaptation
\citep{alvarez2022flowunsteady}. No modification to the underlying viscous-diffusion treatment is introduced in the present work.

\subsection{From External Virtual Work to the Global Nodal Force Vector}
\label{app:work_to_nodal_force}

The structural formulation presented in this work is derived from the principle of virtual work, where the external loading enters through the external virtual work functional

\begin{equation}
\delta W_{\mathrm{ext}}
=
\int_{\Gamma_t}
\mathbf{t}\cdot\delta\mathbf{u},d\Gamma,
\label{eq:app_virtual_work}
\end{equation}

with $\mathbf{t}$ denoting the distributed traction acting on the structural boundary $\Gamma_t$, and $\delta\mathbf{u}$ representing an admissible virtual displacement. Following the finite element discretization, the displacement field and its virtual counterpart are approximated using the interpolation (shape) functions,

\begin{equation}
\mathbf{u}
=
\mathbf{N}\mathbf{d},
\qquad
\delta\mathbf{u}
=
\mathbf{N}\delta\mathbf{d},
\end{equation}

where $\mathbf{N}$ is the matrix of finite element shape functions and $\mathbf{d}$ is the vector of nodal degrees of freedom. Substituting these approximations into Eq.~\eqref{eq:app_virtual_work} gives

\begin{equation}
\delta W_{\mathrm{ext}}
=
\delta\mathbf{d}^{T}
\int_{\Gamma_t}
\mathbf{N}^{T}\mathbf{t},d\Gamma;\quad
\mathbf{F}
=
\int_{\Gamma_t}
\mathbf{N}^{T}\mathbf{t},d\Gamma,
\end{equation}

The integral is identified as the equivalent global nodal force vector \textbf{F}, such that

\begin{equation}
\delta W_{\mathrm{ext}}
=
\delta\mathbf{d}^{T}\mathbf{F}.
\end{equation}

Likewise, the inertia, damping and internal virtual work contributions are discretized to obtain the global mass, damping and stiffness matrices. Consequently, the weak form of the structural equilibrium equations is transformed into the semi-discrete matrix equation

\begin{equation}
\mathbf{M}\ddot{\mathbf{u}}
+
\mathbf{C}\dot{\mathbf{u}}
+
\mathbf{K}\mathbf{u}
=
\mathbf{F}.
\end{equation}

In the present partitioned fluid-structure interaction framework, the equivalent nodal force vector is not assembled by explicitly evaluating Eq.~\eqref{eqn:final_solid_eqn}. Instead, the conservative work force-transfer procedures described in Section. (\ref{sec:work_transfer}) directly provides an energetically equivalent nodal force vector on the structural mesh. Therefore, the transferred aerodynamic loads are incorporated directly into the right-hand side of the semi-discrete structural equations while remaining mathematically equivalent to the virtual work formulation.

\subsection{Mixed Element Representation in FEniCS}
The \textit{Reissner-Mindlin} plate formulation is discretized using a mixed finite
element method based on the Durán-Liberman element, which is specifically
designed to alleviate shear locking in thin plate analyses. The formulation
introduces four field variables: the rotation of the plate normal
$\boldsymbol{\theta}$, the transverse displacement $w$, the reduced shear
strain $\boldsymbol{\gamma}_R$, and a Lagrange multiplier $\mathbf{p}$ that
enforces consistency between the shear strain computed from the primal
variables and the reduced shear strain.

The finite element approximation employs different interpolation orders for
each field. The rotation field is approximated using continuous quadratic
Lagrange vector elements, while the transverse displacement is interpolated
using continuous linear Lagrange elements. Both the reduced shear strain and
the Lagrange multiplier are discretized using first-order Nédélec edge
elements ($NED_1$), resulting in the mixed approximation space

$$
\boldsymbol{\theta}\in [CG_2]^2,\qquad
w\in CG_1,\qquad
\boldsymbol{\gamma}_R\in NED_1,\qquad
\mathbf{p}\in NED_1.
$$

The complete mixed finite element space is therefore given by,

$$
V_h=[CG_2]^2\times CG_1\times NED_1\times NED_1.
$$

Hence, the unknown state vector is

$$
\mathbf{q}=
\left(
\boldsymbol{\theta},
w,
\boldsymbol{\gamma}_R,
\mathbf{p}
\right),
$$

where the first two variables constitute the primal unknowns, whereas
$\boldsymbol{\gamma}_R$ and $\mathbf{p}$ are auxiliary variables introduced to obtain a stable mixed formulation. The reduced shear strain $\boldsymbol{\gamma}_R$ provides an independent approximation of the transverse shear strain, while the Lagrange multiplier weakly enforces the constraint

$$
\boldsymbol{\gamma}
=
\nabla w-\boldsymbol{\theta}
\approx
\boldsymbol{\gamma}_R,
$$

thereby preventing shear locking while preserving the consistency of the Reissner-Mindlin theory. Although the Reissner–Mindlin formulation is discretized using the mixed state vector, only the physical kinematic variables (translations and rotations) are communicated across the fluid–structure interface. The auxiliary reduced shear strain and Lagrange multiplier are internal variables introduced solely to stabilize the finite element discretization and are not transferred during the FSI coupling.



\printcredits

\end{document}